\RequirePackage[2020-02-02]{latexrelease}
\documentclass[twocolumn]{aastex631}

\makeatletter
\global\let\tikz@ensure@dollar@catcode=\relax
\makeatother
\usepackage{urwchancal}

\usepackage{amsmath}
\usepackage{graphicx}
\usepackage{natbib}
\usepackage{scalerel}
\usepackage{makecell}
\usepackage{multirow}
\usepackage{txfonts}
\usepackage{verbatim}
\usepackage{color}

\newcommand\soutm{\bgroup\markoverwith
{\textcolor{black}{\rule[0.5ex]{2pt}{0.8pt}}}\ULon}

\shorttitle{Positive AGN feedback in semi-analytics}
\shortauthors{Contini et al.}

\begin{document}

\title{The Impact of Positive AGN Feedback on the Properties of Galaxies in a Semi-Analytic Model of Galaxy Formation}

\author{Emanuele Contini}
\affil{Department of Astronomy and Yonsei University Observatory, Yonsei University, 50 Yonsei-ro, Seodaemun-gu, Seoul 03722, Republic of Korea}
\email{emanuele.contini82@gmail.com}
\author{Sukyoung K. Yi}
\affil{Department of Astronomy and Yonsei University Observatory, Yonsei University, 50 Yonsei-ro, Seodaemun-gu, Seoul 03722, Republic of Korea}
\email{yi@yonsei.ac.kr}
\author{Seyoung Jeon}
\affil{Department of Astronomy and Yonsei University Observatory, Yonsei University, 50 Yonsei-ro, Seodaemun-gu, Seoul 03722, Republic of Korea}
\author{Jinsu Rhee}
\affil{Department of Astronomy and Yonsei University Observatory, Yonsei University, 50 Yonsei-ro, Seodaemun-gu, Seoul 03722, Republic of Korea}
\affil{Korea Astronomy and Space Science Institute, 776, Daedeokdae-ro, Yuseong-gu, Daejeon 34055, Republic of Korea}

\begin{abstract}
We introduce the state-of-the-art semi-analytic model {\small FEGA} (Formation and Evolution of GAlaxies), which incorporates updated prescriptions for key physical processes in galaxy formation. Notably, {\small FEGA} features an unprecedented semi-analytic modeling of positive Active Galactic Nuclei (AGN) feedback. The model combines the latest prescriptions for gas infall and cooling, a revised star formation recipe that incorporates the extended Kennicutt-Schmidt relation, disk instability, updated supernovae feedback, reincorporation of ejected gas, hot gas stripping from satellite galaxies, and the formation of diffuse light. A novel description of AGN feedback is introduced, describing the positive mode as a burst of star formation from a cooling gas fraction.
{\small FEGA} is rigorously calibrated using an MCMC procedure to match the evolution of the stellar mass function from high redshift to the present. Subsequently, the model is tested against several observed and predicted scaling relations, including the star formation rate-mass, black hole-bulge and stellar mass, stellar-to-halo mass, and red fraction-mass relations. Additionally, we test {\small FEGA} against other galaxy properties such as the distribution of specific star formation rates, stellar metallicity and morphology. Our results demonstrate that the inclusion of positive AGN feedback can co-exist with its negative counterpart without drastic alterations to other prescriptions. Importantly, this inclusion improves the ability of the model to describe the primary scaling relations observed in galaxies.
\end{abstract}


\section{Introduction}
\label{sec:intro}

Galaxy evolution involves a complex interplay of physical processes occurring at various redshifts, across a wide range of halo masses, and over different timescales. Over the years, various theoretical approaches have been developed to tackle the challenging task of understanding the role of baryonic physics in galaxy formation and evolution. Analytic models (\citealt{lin-mohr2004,purcell2007,hansen2009,moster2010,leauthaud2012,birrer2014,lu2015,shankar2017,moster2018,kravtsov2018,behroozi2019,contini2020}), semi-analytic models (\citealt{bower2006,croton2006,delucia2007,somerville2008,fontanot2009,guo2011,henriques2013,lee2013,contini2014,hirschmann2016,cora2018,henriques2020,delucia2024,stevens2024}), and numerical simulations (\citealt{springel2001,teyssier2002,springel2005,angulo2012,hopkins2015,pakmor2016,rantala2017,springel2021}) have been developed to shed light on how, when, and where baryonic physics plays a crucial role in shaping the formation and evolution of galaxies.

In the context of semi-analytic models (hereafter SAMs), significant improvements have been made over the last decade (\citealt{guo2013,henriques2015,hirschmann2016,cora2018,xie2020,henriques2020,stevens2024}). Initially designed to provide a reliable description of galaxy evolution, early SAMs utilized the extended Press-Schechter theory (\citealt{press-schechter1974,lacey-cole1993}) to populate dark matter halos with baryonic content. With the advent of the numerical simulation era, SAMs began to utilize more comprehensive and information-rich merger trees (e.g., \citealt{behroozi2013}) to achieve the same objective (see references above). The capacity of merger trees to retain information not only about halos or subhalos at a specific time, but also their formation and assembly histories represented a significant advancement in semi-analytic modeling.

Merger trees serve as the fundamental input for semi-analytic models (e.g., \citealt{lee2014}). Each SAM allocates a quantity of hot gas to every newly formed halo based on the observed baryon fraction (e.g., Planck Collaboration). Under certain conditions, this hot gas can cool, setting the stage for star formation (\citealt{white-rees1978}). Once stars are formed, they can trigger supernova explosions, which reheat some of the gas and expel a portion of it from the halo into an ejected reservoir (e.g., \citealt{henriques2019}). This ejected material may be reaccreted at later stages, initiating the cycle anew. As galaxies form and evolve, they can undergo mergers (\citealt{delucia2007}), which contribute to the growth of the supermassive black hole (BH) located at their center. BHs also grow through the accretion of surrounding gas and exhibit feedback mechanisms by ejecting energy into the interstellar medium. This process can decelerate or even halt cooling (e.g., \citealt{kravtsov-borgani2012} and references therein). These various processes are incorporated into modern SAMs, each with its own set of prescriptions, all aimed at accurately describing the primary properties of galaxies.

Current SAMs incorporate only the negative mode of AGN feedback (see, e.g., \citealt{croton2006,bower2006} for pioneering attempts to model it). In this mode, the central AGN accretes gas from its surroundings. Although this mechanism of BH growth is not particularly efficient, the AGN responds by injecting mechanical energy into the medium through collimated jets or winds (e.g., \citealt{croton2006,guo2011}). These jets or winds can counteract the cooling of the gas, with the implicit assumption that this occurs uniformly throughout the halo. Typically, this feedback mode plays a crucial role in halos with masses greater than $\log M_{\text{halo}} \sim 13$ (\citealt{guo2011,silk2013,henriques2019,fontanot2020}). As the host galaxy grows, the AGN regulates the cooling process, preventing excessive growth (\citealt{monaco2007,lagos2008,somerville2008,kravtsov-borgani2012,vogelsberger2014,schaye2015,zinger2020,bluck2023}). The power of the BH is governed by the efficiency of gas accretion, a crucial yet poorly constrained parameter that significantly influences the high-mass end of the stellar mass function at low redshifts. However, this mode of AGN feedback is notably less effective at higher redshifts (\citealt{silk2024}).

A missing component in SAMs that contributes to further star formation is the positive mode of AGN feedback. A growing body of observational evidence (\citealt{zinn2013,salome2015,cresci2015a,cresci2015b,mahoro2017,salome2017,shin2019,nesvadba2020,joseph2022,tamhane2022,venturi2023,gim2024}, among others) suggests that AGN activity can actually promote star formation rather than merely offsetting cooling and preventing the formation of new stars. Theoretical frameworks have been proposed to address this phenomenon (e.g., \citealt{silk2024} and references therein), and numerical simulations specifically aimed at investigating this aspect have been developed in recent years (e.g., \citealt{gaibler2012,zubovas2013,zubovas2017,mukherjee2018,mercedes-feliz2024}). In the era of increasingly sophisticated SAMs, it becomes clear that for a comprehensive understanding of the key processes governing galaxy formation and evolution, the positive mode of AGN feedback cannot be overlooked. Therefore, modern SAMs should either incorporate a description of this mode or lay the groundwork for achieving that objective. This work aims to address this gap.

To achieve our objective, we have developed the semi-analytic model {\small FEGA} (Formation and Evolution of GAlaxies), which incorporates state-of-the-art implementations of the most significant physical processes that play a crucial role in galaxy formation. Two particular innovations introduced in {\small FEGA} stand out: a revised prescription for star formation that considers the presence of newly formed stars when calculating the amount of cold gas converted into stars, and a novel recipe that incorporates the positive mode of AGN feedback. In the subsequent sections of this paper, we demonstrate that the updated star formation prescription results in an efficiency that increases with stellar mass, consistent with recent observations (e.g., \citealt{shi2018}). Additionally, the new AGN feedback recipe not only addresses an observed process that is absent in contemporary SAMs, but also aligns well with the evolution of the stellar mass function and overall galaxy properties.

In Section \ref{sec:model}, we offer an in-depth description of the key processes incorporated into {\small FEGA}, specifically: gas infall (Section \ref{sec:infall}), gas cooling (Section \ref{sec:cooling}), star formation (Section \ref{sec:starformation}), disk instability (Section \ref{sec:di_inst}), gas stripping (Section \ref{sec:gasstripping}), supernovae feedback (Section \ref{sec:SNfeedback}), reincorporation of ejected gas (Section \ref{sec:reincorporation}), and AGN feedback (Section \ref{sec:agnfeedback}), which encompasses both negative and positive modes. Since some model parameters require tuning, Section \ref{sec:calibration} details our calibration approach. In Section \ref{sec:results}, we conduct an exhaustive analysis of key relationships among galaxy properties, including the star formation rate (SFR) versus stellar mass, BH-bulge/stellar mass, stellar-to-halo mass ratio versus halo mass, and red fraction versus mass. Additionally, we investigate the distribution of specific star formation rates (SSFR) and galaxy morphology as functions of stellar mass. We conclude this analysis by comparing the contributions of the positive mode of AGN feedback to central and satellite galaxies separately. Section \ref{sec:discussion} is devoted to discussing the main findings of our analysis, and finally, Section \ref{sec:conclusion} summarizes our key conclusions. We adopt a Chabrier (\citealt{chabrier2003}) initial mass function for calculating stellar masses, and unless otherwise specified, all units are h-corrected.

\section[]{FEGA}
\label{sec:model}

{\small FEGA} was originally designed and developed with the primary objective of providing an accurate description of the formation and evolution of the intracluster light (ICL). To achieve this, several critical requirements must be met. In today's context, every SAM must offer a robust description of the stellar mass function (SMF) evolution, and predictions for various galaxy properties should closely align with observational data. Within the framework of ICL formation, the evolution of the SMF holds particular significance, given that a significant portion of the diffuse light originates from stellar stripping of satellite galaxies. Consequently, inaccuracies in predicting the galaxy number density could introduce biases in the estimated ICL content (\citealt{contini2014}). As highlighted in the Introduction, a common limitation of many SAMs is the absence of a prescription for the positive mode of AGN feedback, despite this being an observed phenomenon (see references above).

For these reasons, {\small FEGA} is designed to provide a comprehensive representation of the most critical galaxy properties, beginning with the evolution of the SMF, and incorporating a prescription for the positive mode of AGN feedback. In the subsequent sections, we provide a detailed description of the prescriptions adopted by {\small FEGA} for key processes in galaxy formation, including gas infall and cooling, star formation, gas stripping, supernovae (SN) feedback, reincorporation of ejected gas, and notably, a novel AGN feedback prescription that encompasses both negative and positive modes. We conclude this section by briefly outlining our calibration methodology.

\subsection[]{Infall of Gas}
\label{sec:infall}

Galaxies are initialized by filling dark matter halos with a quantity of gas set to match the observed baryon fraction (e.g., Planck Collaboration). Moreover, at each time step, the infall of gas is regulated to ensure that the baryonic mass within halos always corresponds to the baryon fraction. As halos grow, primordial gas is continuously added in proportion, occupying the hot gas reservoir that contains gas capable of cooling (see Section \ref{sec:cooling}).

However, it is well-established that the measured baryon fraction is lower in the halos of dwarf galaxies (\citealt{guo2010}). To account for this reduced efficiency on these halo mass scales, possibly due to photoheating by a UV background (discussed in \citealt{hirschmann2016}), a formula originally proposed by \cite{gnedin2000} describing the dependence of the baryon fraction on halo mass and redshift is implemented in {\small FEGA}. At each time and for a given halo mass, the new baryon fraction is calculated as:
\begin{equation}\label{eqn:modifier}
f_{\rm{b}} (M_{200},z) = \frac{f_{\rm{b}}^{\rm{cosmic}}}{[1 + 0.26M_F (z)/M_{200}]^3}
\end{equation}
Here, the filtering mass $M_F (z)$ determines the range where the baryon fraction is effectively reduced and is calculated following \cite{kravtsov2004}. For halo masses $M_{\rm{200}}>>M_F (z)$, the baryon fraction matches the universal value, while for halo masses close to $M_F (z)$, the reduction in baryon fraction is minimal. For masses significantly smaller than the filtering mass, the baryon fraction can be substantially reduced. Alternatively, {\small FEGA} can compute the new baryon fraction similarly to the model by \cite{guo2011}, where they employ a modified version of Equation \ref{eqn:modifier}, with the filtering mass calculated based on the numerical findings of \cite{okamoto2008}.

\subsection[]{Cooling of Gas}
\label{sec:cooling}

Cooling of gas is a crucial prerequisite for star formation, providing galaxies with the necessary material to form stars. This essential prescription is generally treated in a standard manner in SAMs, with slight variations in some models (e.g., \citealt{stevens2024}). When infalling gas joins a dark matter halo, it undergoes shock heating. If the infall occurs at early times and in halos with masses below $\sim 10^{12} M_{\odot}$ (\citealt{white-rees1978}), which seems to represent a sort of threshold, the shock occurs close to the central regions, allowing the gas to reach the disk at the free-fall rate. Conversely, for more massive halos above this threshold and at later times, the shocks are further from the central regions (near the virial radius), and the shock-heated gas forms a quasi-static hot sphere. At this stage, the gas is assumed to flow gradually towards the central regions through cooling flows.

The first phase described above is termed \emph{rapid cooling} or \emph{infall-dominated regime}, while the second phase is known as the \emph{hot phase regime}. To distinguish between these two phases and enable gas cooling, we need to determine the cooling time $t_{\rm{cool}}$ and the cooling radius $r_{\rm{cool}}$ at each time step. Following \cite{springel2001}, we calculate the cooling rate of the gas in the hot phase regime by assuming that the gas is shock-heated to the virial temperature. The cooling time $t_{\rm{cool}}$ is then given by:
\begin{equation}\label{eqn:tcool}
t_{\rm{cool}}(r) = \frac{3\mu m_H k T_{200}}{2\rho_{\rm{hot}}(r)\Lambda (T_{\rm{hot}}, Z_{\rm{hot}})}.
\end{equation}
Here, $\mu m_{H}$ is the mean particle mass, $k$ is the Boltzmann constant, $T_{200}$ is the virial temperature of the host halo given by $T_{200} = 35.9(V_{200} /\rm{km s^{-1}})^2$, and $\Lambda (T_{\rm{hot}}, Z_{\rm{hot}})$ represents the cooling functions (\citealt{sutherland1993}), dependent on the hot gas temperature, $T_{\rm{hot}}$, and metallicity, $Z_{\rm{hot}}$. Assuming an isothermal sphere for the hot gas density as a function of radius, and equating the cooling time to the halo dynamical time, we can determine the cooling radius, given by:
\begin{equation}\label{rcool}
r_{\rm{cool}} = \left(\frac{t_{\rm{dyn}}M_{\rm{hot}}\Lambda (T_{\rm{hot}}, Z_{\rm{hot}})}{6\pi \mu m_H k T_{200}R_{200}}\right)^{1/2},
\end{equation}
where $t_{\rm{dyn}}$ represents the halo dynamical time defined as the ratio between the virial radius $R_{200}$ and the virial velocity $V_{200}$, and $M_{\rm{hot}}$ is the available mass of the gas.

The two regimes described above are differentiated by comparing the cooling and virial radii. If $r_{\rm{cool}}>R_{200}$, the halo is in rapid cooling, and the hot gas is accreted in free-fall, given by:
\begin{equation}\label{eqn:mcooling1}
\dot{M}_{\rm{cool}} = \frac{M_{\rm{hot}}}{t_{\rm{dyn}}}.
\end{equation}
Conversely, if $r_{\rm{cool}}<R_{200}$, the halo is assumed to be in the hot phase regime, and the cooling flow onto the galaxy is expressed as:
\begin{equation}\label{eqn:mcooling2}
\dot{M}_{\rm{cool}} = M_{\rm{hot}}\frac{r_{\rm{cool}}}{R_{200}t_{\rm{dyn}}}.
\end{equation}
These equations govern the amount of gas available for cooling and, consequently, potential star formation at any given time. In the following sections, we describe our approach to modeling star formation, offering a revised prescription compared to previous models.

\subsection[]{Star Formation}
\label{sec:starformation}

Star formation is a fundamental process in galaxy formation, typically modeled in SAMs using a simplified version of the Kennicutt relation (\citealt{kennicutt1998}), which links the star formation rate density with the gas density. Stars are assumed to form efficiently in disk regions where the surface mass density exceeds a critical value. Following \cite{croton2006}, this critical surface density (also suggested by \citealt{kauffmann1999}) is translated into a critical mass density as follows:
\begin{equation}
M_{\rm{crit}} = 3.8\cdot10^9 \left(\frac{V_{\rm{max}}}{200\,\rm{km/s}}\right)\left(\frac{R_{\rm{gas,d}}}{10\,\rm{kpc}}\right) M_{\odot},
\end{equation}
where $V_{\rm{max}}$ is the maximum circular velocity of the host halo, and $R_{\rm{gas,d}}$ is the scale radius of the gas disk. The stellar and gas disk radii, as well as the radius of the bulge, are computed as in \cite{guo2011}.

The rate at which cold gas is converted into stars is given by the following equation:
\begin{equation}\label{eqn:starform}
\dot{M}_* = \alpha_{\rm{SF}}\frac{M_{\rm{cold}}-M_{\rm{crit}}}{t_{\rm{dyn}}},
\end{equation}
where $\alpha_{\rm{SF}}$ represents the efficiency of star formation, $M_{\rm{cold}}$ is the available mass in cold gas, and $t_{\rm{dyn}} = 3R_{\rm{gas,d}}/V_{\rm{max}}$ is the dynamical time of the disk. To account for the short lifetimes of massive stars, a fraction of the mass given by Equation \ref{eqn:starform} is immediately returned to the cold gas. Assuming a \cite{chabrier2003} initial mass function, the adopted fraction is $R=0.43$. The efficiency of star formation $\alpha_{\rm{SF}}$ is a free parameter that varies from one model to another.

A novel feature of {\small FEGA} regarding this prescription is the assumption that star formation follows the so-called extended Kennicutt-Schmidt relation (EKS) (\citealt{shi2011}), which considers the role played by pre-existing stars. \cite{shi2011} demonstrated that the KS relation can be reformulated by adding the stellar mass surface density, and the exponent in the relation is not zero, as one would expect in the KS relation. Furthermore, they showed that the star formation efficiency is directly related to the stellar mass surface density, implying a higher efficiency for higher stellar mass surface densities (see also \citealt{shi2018}). Under the assumption of an EKS relation, we model the star formation efficiency, $\alpha_{\rm{SF}}$, using the following equation:
\begin{equation}\label{eqn:newsfreff}
\alpha_{\rm{SF}} = a_{\rm{SF}}\log M_* + b_{\rm{SF}},
\end{equation}
where $M_*$ represents the amount of stars already formed, and $a_{\rm{SF}}$ and $b_{\rm{SF}}$ are the slope and intercept of the relation, respectively. These two parameters are determined during the model calibration (see Section \ref{sec:calibration}), with the expected outcome being a higher efficiency of star formation in more massive galaxies. A fixed fraction of metals per unit stellar mass is assumed, which is set post-calibration to match the observed high-mass end of the stellar mass-metallicity relation at the present epoch and is considered valid at all redshifts.

\subsection[]{Disk Instability}
\label{sec:di_inst}

The instability of the disk, apart from mergers, is a significant channel for bulge growth (\citealt{guo2011,tonini2016,lagos2018,irodotou2019,izquierdo-villalba2019,henriques2020,stevens2024} and references therein). Under certain conditions, disk self-gravity can be the dominant factor leading to disk instability. In SAMs, the disk instability is typically addressed using the criterion proposed by \cite{efstathiou1982}, where the disk becomes unstable if
\begin{equation}\label{eqn:di_crit}
V_{\rm{max}} < \sqrt{\frac{GM_{*,\rm{disk}}}{R_{*,\rm{disk}}}},
\end{equation}
with $M_{*,\rm{disk}}$ and $R_{*,\rm{disk}}$ denoting the stellar mass and exponential scale length of the stellar disk, respectively. Upon the disk becoming unstable, a certain excess of stellar mass is assumed to transition from the disk to the bulge. This mass is the difference between the stellar mass of the disk and the critical value that can be derived from Equation \ref{eqn:di_crit}. We model the critical mass of the disk using a revised version of the commonly employed formula:
\begin{equation}\label{eqn:di_mcrit}
M_{*,\rm{crit}} = \delta_{\rm{DI}}\frac{V_{\rm{max}}^2 R_{*,\rm{disk}}}{G},
\end{equation}
where the free parameter $\delta_{\rm{DI}}$ is determined during the model calibration. While earlier SAMs largely adhered to the results of 2D simulations by \cite{mo1998}, where $\delta_{\rm{DI}}$ is 1, more recent SAMs (e.g., \citealt{lagos2018}) or updated versions of earlier SAMs (e.g., \citealt{izquierdo-villalba2019}) either assume or find values for $\delta_{\rm{DI}}$ that deviate significantly from 1.

Given the findings from the studies cited above, this revised disk instability criterion is anticipated to more accurately reproduce galaxy morphologies than the classic one ($\delta_{\rm{DI}}=1$), a point we will thoroughly discuss in Section \ref{sec:overview}.

\subsection[]{SN Feedback}
\label{sec:SNfeedback}

Feedback from massive stars that explode as supernovae is crucial to galaxy formation, and it plays a central role in this model. SN feedback is known to shape the SMF or luminosity function at the low-mass end (e.g., \citealt{guo2010,henriques2013}), preventing dwarf galaxies from excessive growth. The radiative and mechanical energy released during these explosions can profoundly alter the interstellar medium (ISM) by ionizing and heating it. In extreme cases, this heated gas may even be expelled from the halo, a phenomenon integral to all SAMs of galaxy formation. Historically, SN feedback has been employed to its fullest extent because it is the only mechanism enabling models to align with the observed evolution of the low-mass end of the SMF (see, e.g., \citealt{guo2011,guo2013,henriques2013,henriques2015,hirschmann2016,henriques2020} and references therein).

In {\small FEGA}, SN feedback is modeled to heat cold gas in the disk, transferring it to the hot phase, and to shift hot gas to the ejecta reservoir. This feedback is formally represented as in \cite{guo2011}, albeit with different parameter values. Given its pivotal role in shaping the SMF evolution over time—the primary calibration goal—we adopt two parameter sets. One mirrors that of \citealt{henriques2020}, potent since high redshifts. The other, from \cite{hirschmann2016}, features a redshift-dependent SN feedback, becoming less potent over time.

The \cite{guo2011} model calculates the amount of cold gas heated by SN explosions and subsequently injected into the hot component using:
\begin{equation}
\delta M_{\rm{reh}} = \epsilon_{\rm{reh}} \cdot \left[0.5+\frac{V_{\rm{max}}}{V_{\rm{reh}}} \right]^{-\beta_{\rm{reh}}}\cdot \delta M_* ,
\end{equation}
where $\epsilon_{\rm{reh}}$, $V_{\rm{reh}}$, and $\beta_{\rm{reh}}$ are free parameters, and $\delta M_*$ represents the mass of newly formed stars. The inclusion of the variable $\epsilon_{\rm{reh}}$ by \cite{guo2011} allows for enhanced ejection efficiencies in dwarf galaxies. The total energy injected into the disk and halo gas is:
\begin{equation}
\delta E_{\rm{SN}} = \eta_{\rm{ej}} \cdot \left[0.5+\frac{V_{\rm{max}}}{V_{\rm{ej}}} \right]^{-\beta_{\rm{ej}}}\cdot \delta M_* \cdot 0.5V_{\rm{SN}}^2,
\end{equation}
where $\eta_{\rm{ej}}$, $V_{\rm{ej}}$, and $\beta_{\rm{ej}}$ are free parameters, and $0.5V_{\rm{SN}}^2$ represents the average kinetic energy of SN ejecta per unit mass of stars formed, calculated based on $V_{\rm{SN}}=630$ km/s (\citealt{croton2006}). From this energy amount, the ejected gas (including metals) from the halo is:
\begin{equation}
\delta M_{\rm{ej}} = \frac{\delta E_{\rm{SN}}-0.5\delta M_{\rm{reh}}V_{200}^2}{0.5V_{200}^2}.
\end{equation}
The ejected material populates the ejecta reservoir and may be reincorporated at later stages (see Section \ref{sec:reincorporation}). As elaborated further, {\small FEGA} employs two distinct prescriptions for reincorporation time: one with consistently strong SN feedback and another with a more redshift-dependent, gentler SN feedback. Henriques et al.'s studies demonstrate that potent SN feedback should align with a lengthy reincorporation time, with reincorporated material shaping the SMF near its knee at lower redshifts.

\subsection[]{Gas Stripping}
\label{sec:gasstripping}

The stripping of hot gas within the dark matter subhalo of satellite galaxies can be substantial under certain conditions. Ram-pressure stripping is a well-documented phenomenon observed in numerous clusters (e.g., \citealt{gunn1972,abadi1999}), playing a pivotal role in altering galaxy colors. When deprived of cold gas, essential for fueling star formation, galaxies tend to turn redder and passive. In earlier models (e.g., \citealt{delucia2007}), hot gas stripping was considered instantaneous upon a galaxy becoming a satellite. However, contemporary models employ a more gradual approach to gas stripping, enabling satellites to retain a reservoir of hot gas even after crossing the virialized region of halos. Most models link gas stripping to dark matter stripping (e.g., \citealt{guo2011,hirschmann2016,henriques2020}), and {\small FEGA} is no exception.

We model gas stripping by calculating two characteristic radii: the tidal radius, associated with gradual dark matter particle stripping, and the ram-pressure stripping radius, linked to ram-pressure forces exerted by satellite motion through the ISM. The smaller of these two radii dictates the amount of gas stripped. The tidal radius is calculated as:
\begin{equation}
R_{\rm{tid}} = \left(\frac{M_{\rm{dm}}}{M_{\rm{dm,infall}}}\right)\cdot R_{\rm{dm,infall}},
\end{equation}
where the fraction $M_{\rm{dm}}/M_{\rm{dm,infall}}$ represents the lost dark matter amount, and $R_{\rm{dm,infall}}$ denotes the virial radius of the subhalo at its last central state before becoming a satellite.

The ram-pressure force-induced radius, $R_{\rm{ram}}$, is calculated by equating self-gravity of the satellite with the ram-pressure force:
\begin{equation}
\rho_{\rm{\rm{sat}}}(R_{\rm{ram}})V_{\rm{\rm{sat}}}^2 = \rho_{\rm{hot}}(R)V_{\rm{orb}}^2 ,
\end{equation}
where $\rho_{\rm{\rm{sat}}}(R_{\rm{ram}})$ represents the hot gas density of the satellite at $r=R_{\rm{ram}}$, $V_{\rm{\rm{sat}}}$ is the virial velocity of the subhalo calculated at infall, $\rho_{\rm{hot}}$ is the hot gas density of the parent halo at a distance $R$ from the centre of its potential well, and $V_{\rm{orb}}$ stands for the orbital velocity of the satellite, estimated as the virial velocity of the parent halo.

The hot gas of the satellite is assumed to be distributed isothermally, similar to the cooling prescription. All hot gas beyond $r > \min(R_{\rm{ram}},R_{\rm{tid}})$ is stripped and incorporated into the hot gas of the central galaxy. Unlike previous models (e.g., \citealt{guo2011}), which also assume that the same fraction of ejected material is stripped, {\small FEGA} posits that all ejected mass is completely stripped during the initial stripping episode. This assumption is grounded in the idea that the ejected material is likely situated further from the subhalo than where the hot gas is distributed. Moreover, considering the distinction between satellites still tethered to a subhalo, termed type 1 satellites, and those deprived, known as type 2 or orphan galaxies, the former are treated as centrals if located outside the virial radius $R_{200}$ and receive stripped material from orphans stripped beyond $R_{200}$.

\subsection[]{Reincorporation of Ejected Gas}
\label{sec:reincorporation}

As mentioned in Section \ref{sec:SNfeedback}, the timescale for the reincorporation of ejected material into the hot component is crucial. Typically, the reincorporation time depends on the host halo-subhalo dynamical time (\citealt{delucia2007,guo2011,hirschmann2016,delucia2024}). However, early SAMs often exhibited an overabundance of low-mass galaxies ($\log M_* < 9.5$) at early times (\citealt{fontanot2009,weinmann2012,guo2013,vogelsberger2014,hirschmann2016}). \cite{henriques2013} addressed this issue by introducing strong stellar feedback in low-mass galaxies coupled with extended reincorporation times $t_{\rm{reinc}}$ for the ejected material. This approach effectively described the evolution of the SMF. Following this work, similar reincorporation times were adopted by other authors (\citealt{hirschmann2016,cora2018}).

However, as pointed out by \cite{hirschmann2016}, there are alternative methods to achieve the same result. One approach is to employ redshift-dependent stellar feedback, becoming gentler over time, but still strong at high redshifts. Another method involves preemptively removing a significant portion of infalling gas in low-mass halos, storing it as ejected gas. This gas could be pre-heated by stellar and/or AGN-driven winds. In this scenario, a longer $t_{\rm{reinc}}$, as in the model by \cite{henriques2013}, is required.

In {\small FEGA}, both reincorporation time prescriptions are implemented. Strong SN feedback, similar to \cite{henriques2020}, is associated with a longer $t_{\rm{reinc}}$, while the redshift-dependent SN feedback is linked to a shorter $t_{\rm{reinc}}$. For the former, $t_{\rm{reinc}}$ is calculated as:
\begin{equation}\label{eqn:tr1}
t_{\rm{reinc}} = \gamma_1 \frac{10^{10}M_{\odot}}{M_{200}}
\end{equation}
where $\gamma_1$ is a free parameter set during calibration, and $M_{200}$ is the virial mass of the halo. The rate of reincorporation of ejected gas is then:
\begin{equation}
\dot{M}_{\rm{ej}} = \frac{M_{\rm{ej}}}{t_{\rm{reinc}}}.
\end{equation}
For the latter, $t_{\rm{reinc}}$ is given by:
\begin{equation}\label{eqn:tr2}
t_{\rm{reinc}} = t_{\rm{dyn}} = \frac{R_{200}}{V_{200}},
\end{equation}
and the rate of reincorporation becomes:
\begin{equation}\label{eqn:rmassA}
\dot{M}_{\rm{ej}} = \gamma_2 \frac{M_{\rm{ej}}}{t_{\rm{reinc}}},
\end{equation}
where $\gamma_2$ is another free parameter determined during calibration.

\subsection[]{AGN Feedback: Negative and Positive}
\label{sec:agnfeedback}

The role of AGN feedback is critical in galaxy formation, balancing the growth of the BH while also regulating the growth of galaxies themselves. Typically, AGN feedback is categorized into two modes: the \emph{quasar mode} and the \emph{radio mode}. The quasar mode primarily describes the BH growth during galaxy mergers, while the radio mode focuses on preventing excessive cooling of the hot gas, thus inhibiting star formation. Here, we focus on the radio mode, presenting two distinct prescriptions: the conventional negative feedback approach and a novel positive feedback mechanism that can enhance star formation under certain conditions. First, let's briefly touch upon the quasar mode.

The growth of the BH is largely attributed to gas-rich mergers. During these mergers, the BH of the primary galaxy grows by assimilating the BHs of the secondary galaxies and a portion of their cold gas. This mechanism is well-established in the history of SAMs, and we adhere to the original formulation proposed by \cite{croton2006}. In this model, the growth in BH mass is expressed as:
\begin{equation}\label{eqn:quasarmode}
\Delta M_{\rm{BH}} = f_{\rm{BH}}\left(\frac{M_{\rm{sat}}}{M_{\rm{cen}}}\right)\left(\frac{M_{\rm{cold}}}{1+(280\, \rm{km/s})/V_{200}}\right),
\end{equation}
where $M_{\rm{sat}}/M_{\rm{cen}}$ represents the ratio of stellar and cold gas masses between the satellite and central galaxies, $M_{\rm{cold}}$ denotes the cold gas mass, and $V_{200}$ is the virial circular velocity of the central halo. The parameter $f_{\rm{BH}}$ modulates the effective amount of cold gas accreted by the final BH. Initially set to 0.03 in \cite{croton2006} to align with the present-day $M_{\rm{BH}}-M_{\rm{bulge}}$ relation, this parameter was later increased to 0.066 in \cite{henriques2020} for models employing the same prescription. We set the value of this parameter to the most updated one provided by Henriques et al. The total BH mass following a merger is thus given by $M_{\rm{BH}}=M_{\rm{BHsat}}+M_{\rm{BHcen}}+\Delta M_{\rm{BH}}$, where $M_{\rm{BHsat}}$ and $M_{\rm{BHcen}}$ are the initial BH masses of the satellite and central galaxies, respectively.

\subsubsection[]{Negative Feedback}
\label{sec:negativefb}

The radio mode of AGN feedback is implemented based on the original version in \cite{croton2006}. This mode accounts for the accretion of hot gas onto the central BH, which subsequently releases energy into the hot gas atmosphere. The gas accretion rate is described by:
\begin{equation}\label{eqn:radiomode}
\dot{M}_{\rm{BH}} = \kappa_{\rm{AGN}} \left(\frac{f_{\rm{hot}}}{0.1}\right)\left(\frac{V_{200}}{200\,  \rm{km/s}}\right)^3 \left(\frac{M_{\rm{BH}}}{10^8 \, M_{\odot}/h}\right)
\end{equation}
in units of $M_{\odot}/\rm{yr}$, where $f_{\rm{hot}}$ represents the ratio of hot gas to dark matter mass, and $\kappa_{\rm{AGN}}$ is a parameter that accounts for the efficiency of accretion. A larger value of $\kappa_{\rm{AGN}}$ indicates more efficient accretion and consequently a larger amount of energy released. We set the value of $\kappa_{\rm{AGN}}$ during the calibration of the model. Following \cite{croton2006}, we assume that the rate of mechanical energy released is:
\begin{equation}
\dot{E}_{\rm{radio}} = \eta_{\rm{rad}} \dot{M}_{\rm{BH}}c^2 ,
\end{equation}
where $\eta_{\rm{rad}}$ is a parameter assumed to be 0.1, and $c$ is the speed of light. This injection of energy counteracts the cooling process (refer to Eq. \ref{eqn:mcooling1} and \ref{eqn:mcooling2}). Therefore, the net rate of gas cooling is given by:
\begin{equation}\label{eqn:mcool}
\dot{M}_{\rm{cool,new}} = \dot{M}_{\rm{cool}}-2\frac{\dot{E}_{\rm{radio}}}{V_{200}^2}.
\end{equation}
The AGN feedback can either entirely halt the cooling of gas or reduce it to some extent. In both scenarios, this marks the end of the negative mode of the AGN feedback. In traditional SAMs, the net mass given by Equation \ref{eqn:mcool} would be added to the cold gas component, making it available for star formation. However, in {\small FEGA}, this amount of cold gas undergoes a second phase of AGN feedback, the positive mode, where it is assumed that a portion of this mass is converted into stars through a burst of star formation.

\subsubsection[]{Positive Feedback}
\label{sec:positivefb}

This prescription represents the second novelty of our model and is an unprecedented addition to semi-analytic approaches. In recent years, accumulating observational evidence (e.g., \citealt{zinn2013,salome2015,cresci2015a,cresci2015b,mahoro2017,salome2017,shin2019,nesvadba2020,joseph2022,tamhane2022,venturi2023,gim2024} and references therein) and numerical tests/theoretical methods (e.g., \citealt{gaibler2012,zubovas2013,silk2013,bieri2015,bieri2016,zubovas2017,mukherjee2018,silk2024} and references therein) have suggested the existence of positive AGN feedback, mainly in radio-loud AGNs, early-phase quasars and low-luminosity AGNs (see references above). In this scenario, the BH promotes star formation rather than suppressing it. In reality, it is quite likely that both negative and positive feedback mechanisms can operate simultaneously. The AGN injects energy into the hot atmosphere either via a jet or wind. While this energy might be sufficient to halt cooling, a crucial question arises: can we be certain that this cooling shutdown process is isotropic and exerts consistent influence throughout the halo?

In the landmark paper by \cite{silk-rees1998}, self-regulated BH growth was proposed to explain the BH mass-velocity dispersion relation, grounding this argument in energy balance. Indeed, as depicted in SAMs, injecting energy into the hot gas atmosphere can deter further accretion onto the BH. However, there is a possibility that the gas near the BH is heated and subsequently cools as it expands. Subsequent studies, such as \cite{silk2010} (but also see \citealt{silk2009,silk2013}), have introduced positive AGN feedback as a solution to this issue.

Drawing inspiration from these theories, we entertain the possibility that a portion of the mass defined by Equation \ref{eqn:mcool} can lead to star formation. Consequently, the positive mode is activated subsequent to the negative one. If the energy released by the BH is sufficient to entirely counterbalance the cooling flow (i.e., $\dot{M}_{\rm{cool,new}} = 0$), the positive mode remains inactive. However, when $\dot{M}_{\rm{cool,new}} > 0$, some of the cold gas undergoes a burst of star formation, which triggers supernovae feedback. Although the precise mechanism through which AGN can enhance star formation remains elusive, we postulate that the cold gas mass is potentially available for star formation. We shape this mass transformation into stars via a functional form, expressed as:
\begin{equation}\label{eqn:pAGN}
\dot{M}_* = \alpha_{\rm{pAGN}}\left(\frac{M_{BH}}{10^{8} \, M_{\odot}/h}\right)^{\beta_{\rm{pAGN}}}\dot{M}_{\rm{cool,new}}.
\end{equation}
Here, $\dot{M}_*$, the rate of newly formed stars, is computed as a function of $\dot{M}_{\rm{cool,new}}$ and is proportional to the BH mass. The parameters $\alpha_{\rm{pAGN}}$ and $\beta_{\rm{pAGN}}$ govern the fraction of mass transformed into stars. Given the current lack of precise knowledge regarding these parameters, we determine them during the model calibration process.

Several caveats warrant discussion. Firstly, the positive mode operates independently of the negative mode. The only circumstance under which this mode is deactivated is when there is an absence of cooling flow, either due to an extremely potent negative mode or no cooling at all. Secondly, this mode is operational across all redshifts without explicit redshift dependence. However, as highlighted in Section \ref{sec:cooling}, cooling is notably efficient at higher redshifts and in low-mass halos, mirroring the efficacy of this mode. Preliminary observations suggest that the parameters $\alpha_{\rm{pAGN}}$ and $\beta_{\rm{pAGN}}$ are configured such that the positive mode exhibits greater effectiveness in less massive halos. This aspect aligns with earlier theories on positive AGN feedback (e.g., \citealt{silk2024} and references therein), which anticipate a more pronounced effect at higher redshifts and in lower mass halos.

\subsection[]{Intracluster Light}
\label{sec:icl}

The ICL stands as a significant observed component within galaxy groups and clusters (\citealt{contini2021,montes2022,contini2024c}). As such, SAMs must offer an accurate depiction of its formation and primary properties, commencing with the observed quantities. Observational data indicate that a notable fraction of stars, ranging from approximately 5\% to 50\% (e.g., \citealt{montes2022}), within the virial radius of halos are unbound to any member galaxy and are solely influenced by the potential well of the host halo. Prior models (e.g., \citealt{purcell2007,guo2011,henriques2013,contini2014,contini2018,contini2019,stevens2024,contini2024a}) and numerical simulations (e.g., \citealt{murante2007,rudick2011,cui2014,tang2018,pillepich2018,montenegro-taborda2023,contreras-santos2024,ahvazi2024}) have aimed to replicate the observed ICL quantities and its properties, including colors across various bands, metallicity, and ages.

Presently, it is established that the diffuse light within clusters originates from multiple sources: tidal stripping of satellite galaxies, the disruption of dwarfs, satellite mergers with the central galaxy, and pre-processing or accretion from outside the halo (refer to \citealt{contini2021} for an exhaustive review). The first two mechanisms are intrinsically linked, stemming from tidal forces exerted on satellites while they orbit the central galaxy, often termed as the brightest group/cluster galaxy (BCG). The third mechanism accounts for the intense relaxation processes during mergers, which give rise to stray stars (e.g., \citealt{murante2007}). Meanwhile, pre-processing denotes the ICL formation elsewhere through the aforementioned processes and its subsequent accretion into the halo during its formation (e.g., \citealt{pillepich2018}). It is worth noting that \emph{in situ} star formation has been proposed previously (\citealt{puchwein2010}) but was largely dismissed as a primary channel due to observational constraints (e.g., \citealt{melnick2012}). However, in a recent study by \cite{ahvazi2024}, they found that between 8\% and 28\% of stars in the ICL originate \emph{in situ} within the TNG50 simulation.

{\small FEGA} incorporates all the processes mentioned above (with the exception of \emph{in situ} star formation), providing a specific prescription for each. In formal terms, we model the formation of the ICL similarly to the approach taken in \cite{contini2014} with the {\small L-Galaxies} model, which was subsequently refined in \cite{contini2018} and \cite{contini2019}. Our model includes prescriptions for stellar stripping due to tidal forces acting on satellites, violent relaxation during mergers, and the accretion of pre-processed material onto a halo.

The stellar stripping of type 2 satellites (orphans without a subhalo) is estimated by calculating the tidal radius $r_{\rm{t}}$ from which tidal forces are sufficiently strong, compared to the gravity of the satellite \footnote{For simplicity, the stellar density profile of satellite galaxies is approximated by a spherically symmetric isothermal profile.}, to strip a certain amount of stars.
The tidal radius $r_{\rm{t}}$ is given by (see \citealt{binney2008}):
\begin{equation}\label{eqn:tid_rad}
r_{\rm{t}} = \left(\frac{M_{\rm{sat}}}{3 \cdot M_{\rm{halo}}}\right)^{1/3} \cdot d,
\end{equation}
where $d$ is the distance of the satellite from the centre of the potential well which, by definition in our model, is occupied by the central galaxy. The satellite is then considered to be a two-component system: a bulge, which is modeled with a Jaffe (\citealt{jaffe1983}) profile, and a disk, modeled with an exponential profile. If $r_{\rm{t}}$ is small enough to be contained within the bulge, the satellite is assumed to be completely destroyed (disruption channel). However, if $r_{\rm{bulge}}<r_{t}<r_{\rm{sat}}$, only the stellar mass within the shell $r_{\rm{t}}-r_{\rm{sat}}$ is stripped. In both cases, the stellar mass content is added to the ICL component of the central galaxy of the halo. In the second case, the scale length $R_{\rm{*,disk}}$ of the new disk of the satellite is updated to $r_{t}/10$ \footnote{Given that we assume an exponential profile for the stellar disk, 99.9\% of the stellar mass is contained within ten times the scale length of the disk.}.

For type 1 satellites, i.e., those that are still associated to a subhalo, the computation of the tidal radius is mathematically given by Equation \ref{eqn:tid_rad}, but the following constraint must be satisfied initially:
\begin{equation}\label{eqn:eq_radii}
 R^{\rm{DM}}_{50} < R^{\rm{Disk}}_{50},
\end{equation}
where $R^{\rm{DM}}_{50}$ is the half-mass radius of the parent subhalo, and $R^{\rm{Disk}}_{50}$ the half-mass radius of the satellite's disk, given $1.68\cdot R_{\rm{sl}}$ for an exponential profile. Once this condition is met, the ICL associated with these satellites is also lost, transferring to the ICL component of the BCG.

The second channel for the formation of ICL in {\small FEGA} is through galaxy mergers, whether minor or major. At the moment of the merger, when the satellite is incorporated by the central galaxy, the model assumes that 20\% of the stellar mass in the satellite becomes unbound and contributes to the ICL component, while the remaining 80\% merges with the central galaxy (\citealt{contini2014}). This particular mechanism for ICL formation will be further refined (Contini et al. in prep.) by introducing a scatter in the fraction of stellar mass that becomes unbound. Additionally, an implementation of the stellar halo of central galaxies will be included, with material from ICL stars returning to the bound phase.

Stellar stripping and mergers are the primary channels for ICL formation in {\small FEGA}. However, pre-processing serves as another significant channel through which halos accumulate ICL (e.g., \citealt{contini2024a} and references therein). This portion of accreted ICL originates from satellite galaxies in two ways: (a) as mentioned earlier, when type 1 satellites undergo initial stripping, their ICL accumulates in the central galaxies; (b) orphan galaxies lose their ICL upon becoming orphans. In \cite{contini2024a}, we demonstrated that this channel accounts for an average of approximately 20\% of the total ICL.

\section[]{Calibration}
\label{sec:calibration}

Calibrating the model is a nuanced endeavor, as SAMs involve the treatment of numerous physical processes with some parameters being either poorly constrained or entirely unknown, such as those listed in Table \ref{tab:parameters}. Various approaches can be employed for calibration. One method is to utilize results from numerical simulations that employ similar treatments for the primary processes involved in galaxy formation (e.g., as done in \citealt{hirschmann2016,delucia2024}). Alternatively, numerical algorithms can be employed to statistically determine the parameter values that best fit a given set of observations, such as the evolution of the SMF and/or luminosity function (e.g., as done in \citealt{cora2018,henriques2020}).

Regardless of the chosen calibration method, the resulting model can only be deemed reliable to a certain degree of accuracy. Furthermore, calibrating a model based on a specific set of observations does not necessarily guarantee accurate predictions for other galaxy properties. Essentially, while calibration is a crucial and essential step, the reliability of the model in predicting specific galaxy properties is ultimately determined by the fidelity of the modeled physics and its implementation.

\subsection{MCMC Approach}\label{sec:mcmc}
The calibration of FEGA is carried out using the Monte Carlo Markov Chains (MCMC) technique, an approach widely employed across various fields of astrophysics and science in general. This method has been successfully used in the past to calibrate the semi-analytic model L-Galaxies (see, for instance, \citealt{henriques2020} and the references therein), establishing it as one of the most reliable SAMs to date. This technique has also found application in estimating cosmological parameters (e.g., \citealt{lewis2002}), analyzing gravitational wave data (e.g., \citealt{veitch2015}), and exoplanet research (e.g., \citealt{foreman-mackey2019}).

The strength of the MCMC approach lies in its ability to replicate various galaxy properties not used during calibration, as demonstrated in the notable works of Henriques et al. The underlying principle of the MCMC approach is straightforward. However, its application can be computationally intensive and somewhat challenging, serving as a double-edged sword if the user lacks a comprehensive understanding of the parameters involved in the calibration and the physics they represent. Essentially, the MCMC approach aims to identify the optimal values for the set of free parameters that align most closely with the observational data the model is intended to replicate within a specified level of accuracy. It is crucial to emphasize that the MCMC approach is purely statistical, devoid of any underlying physics. Consequently, a deep understanding of the significance of the parameters is vital. While the MCMC calibration provides parameter values that closely match observational data, these values may not necessarily hold physical meaning. Instead, they represent the best statistical fit to the data.

The MCMC approach becomes particularly advantageous when dealing with a limited parameter set, as it offers both accuracy and efficiency. The required level of accuracy primarily hinges on the choice of observational data, the number of interactions employed, and, importantly, the merger trees used. The selection of merger trees is critical for two main reasons: (a) the trees must encompass a sufficiently large volume to accurately replicate the observed relations in use, such as stellar mass or luminosity functions; (b) they should include a diverse array of galaxies, representing various morphologies and environments. This diversity is crucial when utilizing relations involving the fraction of passive galaxies or morphological characteristics in the calibration process.

However, computational speed is another vital consideration. While larger merger trees offer more comprehensive data, they also demand significantly higher computational resources. To address this challenge, researchers often opt for truncated sections of the tree. These truncated portions are carefully selected to strike a balance between computational efficiency and the desired level of calibration accuracy, as discussed in the works of Henriques et al.

\subsection{Observations Used in the Calibration}\label{sec:set_obs}

The choice of observational data for calibration in SAMs is a nuanced and critical step. While the temptation is to utilize a broad spectrum of observational data to rigorously constrain the model parameters, this approach can sometimes lead to issues. Over-reliance on observational data can inadvertently bias the model towards fitting specific data points rather than accurately capturing the underlying physical processes governing galaxy formation.

SAMs are constructed to provide a coherent and plausible description of galaxy formation. They aim to encapsulate the key physical processes believed to be pivotal in shaping the observed properties of galaxies. Therefore, a SAM's success should be gauged not only by its ability to fit a chosen set of observational data, but also by its capacity to predict a diverse range of galaxy properties that were not explicitly used in its calibration.

When a wide array of observational data is employed, particularly when the model parameters are numerous, it can overly constrain the model. This approach might force the model to conform too closely to the observed data, potentially at the expense of its ability to accurately represent the complex and multifaceted nature of galaxy formation.

A more prudent approach, as advocated in some seminal works such as those by Henriques et al. and \cite{cora2018}, involves using a limited subset of observational relations for calibration (see also the discussion in \citealt{knebe2018}). By doing so, the model is given the freedom to predict other galaxy properties that were not part of the calibration process. This approach ensures that the predictions of the model are rooted in the broader physical principles of galaxy formation rather than being solely dictated by data-fitting. While observational data are indispensable for calibrating SAMs, the selection and utilization of these data should be done judiciously. A balanced approach that combines a limited set of observational data with the predictive capabilities of the model can yield a more robust and insightful model of galaxy formation.

The calibration of {\small FEGA} is carefully tailored to leverage specific observational data that are deemed critical for constraining its parameters effectively. Given the focused nature of the parameters under consideration, a targeted approach is adopted to ensure the model's predictive accuracy while maintaining its flexibility and robustness.
{\small FEGA} is calibrated solely by using the evolution of the SMF from redshift $z=3$ to the present epoch, and the datasets utilized in the calibration are consistent with those employed in \cite{henriques2015}. These datasets amalgamate contributions from multiple studies (listed below) to provide a comprehensive view of the SMF across various redshifts, specifically $z=3,2.75,2,1.75,1,0.75,0.4,0$. As for reference, the SMFs at $z>0$ are derived from the studies of \cite{marchesini2009}, \cite{marchesini2010}, \cite{ilbert2010} \cite{sanchez2012}, \cite{muzzin2013}, \cite{ilbert2013} and \cite{tomczak2014}. Finally, at the present epoch ($z=0$), the SMF is derived by studies from \cite{baldry2008}, \cite{li-white2009}, \cite{baldry2012} and \cite{bernardi2018}.

\subsection{Simulations Used}\label{sec:simulations}
The final component crucial to the calibration and operation of {\small FEGA} is the suite of cosmological simulations upon which it is based. These simulations not only serve as the foundation for the calibration of {\small FEGA}, but also act as the computational environment in which the model operates to generate its final galaxy catalogs.

The merger trees required for the calibration and the subsequent parameter tuning of {\small FEGA} are obtained from a set of Dark Matter (DM)-only cosmological simulations. Specifically, two simulations from the dataset detailed in \cite{contini2023} are employed for this purpose. The first simulation, YS50HR, covers a cosmological volume of $(50 \,\rm{Mpc/h})^3$ and is primarily utilized for the initial calibration phase. Subsequently, a larger simulation volume, YS200, encompassing $(200 \,\rm{Mpc/h})^3$, is employed for a second round of calibration and to construct the final galaxy catalogs.

Both of these simulations are conducted over a redshift range from $z=63$ to the present epoch ($z=0$). The simulation data is captured in 100 discrete snapshots spanning from $z=20$ to $z=0$. The simulations are executed using {\small GADGET4} (\citealt{springel2021}), a state-of-the-art code for cosmological simulations. The cosmological parameters adopted in these simulations adhere to the Planck 2018 cosmology (\citealt{planck2020}), with the following specific values: $\Omega_m=0.31$ representing the total matter density, $\Omega_{\Lambda}=0.69$ denoting the cosmological constant, $n_s=0.97$ indicating the primordial spectral index, $\sigma_8=0.81$ specifying the power spectrum normalization, and $h=0.68$ representing the normalized Hubble parameter.

In summary, the suite of cosmological simulations forms an integral part of {\small FEGA}, providing the necessary merger trees and a computational framework within which the model is calibrated and operates to generate its final galaxy catalogs.

The YS50HR simulation serves as the foundational dataset for the calibration of {\small FEGA}, offering a mass resolution of $10^7 \,M_{\odot}/h$. Given its resolution and size, which align with the criteria previously discussed, this volume is deemed appropriate for the initial calibration of the model.

In line with the methodology employed in \cite{henriques2013}, the full merger trees from the YS50HR simulation are partitioned to identify a sufficiently large subset of merger trees. This subset is chosen to ensure both a reasonably rapid convergence of results and a representative sampling of the entire cosmological volume.

Following the initial calibration using YS50HR, a second phase of calibration is conducted using a subset of the merger trees from the YS200 simulation. While YS200 offers a volume approximately ten times larger than YS50HR, it is also around ten times less resolved in terms of mass. The primary objective of this subsequent calibration step is to verify the convergence of the results across a broader volume, albeit with reduced mass resolution. This step essentially serves as a refinement to the initial calibration process.

Upon determining the optimal values for the free parameters through this calibration process (refer to Section \ref{sec:par_set}), {\small FEGA} is subsequently executed on the complete merger tree dataset of YS200. This execution enables the generation of the final galaxy catalogs that represent the predictions of the calibrated {\small FEGA} within the simulated cosmological volume.

\subsection{Set of Parameters}\label{sec:par_set}
In the calibration process of {\small FEGA}, a set of eight parameters plays a crucial role. These parameters are instrumental in governing various processes of galaxy formation, including star formation, reincorporation of ejected gas, and both modes of radio mode AGN feedback. Specifically, the parameters are: slope, $a_{\rm{SF}}$, and intercept, $b_{\rm{SF}}$ in Equation \ref{eqn:newsfreff}, the free parameter $\delta_{\rm{DI}}$ in the critical mass of the disk, the free parameters $\gamma_1$ and $\gamma_2$ for the reincorporated gas, the efficiency $\kappa_{\rm{AGN}}$ of the BH in Equation \ref{eqn:radiomode}, and the two parameters $\alpha_{\rm{pAGN}}$ and $\beta_{\rm{pAGN}}$ in Equation \ref{eqn:pAGN}. The list and respective values of these parameters can be found in Table \ref{tab:parameters}.

In the development of {\small FEGA}, we introduce two distinct versions, which are delineated based on their treatments of SN feedback mechanisms, as outlined in Section \ref{sec:SNfeedback}.
The first version, termed ModA, adopts a nuanced and redshift-dependent SN feedback mechanism, drawing inspiration from the model proposed in \cite{hirschmann2016}. This version is associated with the reincorporation time outlined in Equation \ref{eqn:tr2}. Conversely, the second version, denoted as ModB, incorporates a robust and consistent SN feedback model, akin to the approach presented in \cite{henriques2020}. This version is characterized by its coupling with the reincorporation time specified by Equation \ref{eqn:tr1}.

The real differences between the two models lie exclusively on the stellar feedback, the reincorporation of the ejected gas and, of course, to their separate calibrations. These variations allow for an in-depth exploration and understanding of the diverse feedback mechanisms and their implications for galaxy formation within the {\small FEGA} framework.

To summarize, ModA has a SN feedback described by
\begin{equation}
 \delta M_{\rm{reh}} = 0.7 \cdot \left[0.5+(1+z)^3 \frac{V_{\rm{max}}}{70\, \rm{km/s}} \right]^{-3.5}\cdot \delta M_* \, ,
\end{equation}
and
\begin{equation}
 \delta E_{\rm{SN}} = 0.15 \cdot \left[0.5+(1+z)^3 \frac{V_{\rm{max}}}{70\, \rm{km/s}} \right]^{-3.5}\cdot \delta M_* \cdot 0.5V_{\rm{SN}}^2\, ,
\end{equation}
while the reincorporated gas is given by Equation \ref{eqn:rmassA}. ModB, instead, has a SN feedback described by
\begin{equation}
 \delta M_{\rm{reh}} = 5.6 \cdot \left[0.5+\frac{V_{\rm{max}}}{110\, \rm{km/s}} \right]^{-2.0}\cdot \delta M_* \, ,
\end{equation}
and
\begin{equation}
 \delta E_{\rm{SN}} = 5.5 \cdot \left[0.5+\frac{V_{\rm{max}}}{220\, \rm{km/s}} \right]^{-2.9}\cdot \delta M_* \cdot 0.5V_{\rm{SN}}^2\, ,
\end{equation}
while the reincorporation time is given by Equation \ref{eqn:tr1}.

\begin{table}
\begin{center}
  \caption{Main parameters of the models and their values. The first column indicates the name of the parameters, the second and third their values for ModA and ModB, respectively. The fourth column indicates the unit of the parameters, if any, while the fifth and last column states whether the value of the parameter in question was set during the calibration, or fixed.}
\label{tab:parameters}
\begin{tabular}{lccccc}
\hline
\hline
&Parameter &ModA &ModB &units   &fixed/MCMC \\
\hline
&$a_{\rm{SF}}$ &0.0030   & 0.0028    &  -   & MCMC \\
&$b_{\rm{SF}}$ &-0.005  & -0.004   &  -   & MCMC \\
\hline
&$\delta_{\rm{DI}}$ &  0.20 & 0.22 &  -   & MCMC \\
\hline
&$\epsilon_{\rm{reh}}$  &0.7 & 5.6 &  -    & fixed \\
&$\eta_{\rm{ej}}$    & 0.15  & 5.5 &  -    & fixed \\
&$\beta_{\rm{reh}}$  & 3.5  & 2.9 &   -   & fixed \\
&$\beta_{\rm{ej}}$   & 3.5  & 2.0 &   -   & fixed \\
&$V_{\rm{reh}}$  & 70  & 110 & km/s     & fixed \\
&$V_{\rm{ej}}$   & 70  & 220 & km/s     & fixed \\
&$V_{\rm{SN}}$   & 630 & 630 & km/s     & fixed \\
\hline
&$\gamma_{\rm{1}}$   & -  & 1.15 $\cdot 10^{10}$ &   $\rm{yr}$   & MCMC \\
&$\gamma_{\rm{2}}$   & 0.375  & - &  -    & MCMC \\
\hline
&$f_{\rm{\rm{BH}}}$  &0.066 &  0.066   &  -   & fixed \\
&$k_{\rm{AGN}}$ &5e-5  & 1e-5     &  $M_{\odot}$/yr   & MCMC \\
&$\eta_{\rm{rad}}$    & 0.1  & 0.1 &  - & fixed \\
&$\alpha_{\rm{pAGN}}$ &0.032   & 0.007  &  -   & MCMC \\
&$\beta_{\rm{pAGN}}$  &-0.05   & -0.015   &  -   & MCMC \\
\hline
&$f_{\rm{metals}}$  & 0.025   & 0.05   &  -   & fixed \\
\hline
\hline
 \end{tabular}
\end{center}
\end{table}

In the following section, we investigate into the key findings from our comprehensive analysis, encompassing a diverse range of galaxy properties, including both central and satellite galaxies. Additionally, we will present an isolated examination of the positive AGN feedback to quantify its significance across galaxies of varying masses.

Upon examining Table \ref{tab:parameters}, it becomes evident that for both ModA and ModB models, the potency of the positive mode of the AGN feedback exhibits a decreasing trend with increasing BH mass. Considering that more massive haloes normally host more massive BHs, this observation is consistent with our expectations outlined in Section \ref{sec:positivefb}. Concurrently, the efficiency of star formation manifests as an escalating function relative to stellar mass. These insights illuminate the intricate interplay between AGN feedback mechanisms and star formation processes across the galaxy population, underscoring the nuanced dependencies on stellar mass.

\section{Results}
\label{sec:results}

The analysis encompasses a broad spectrum of galaxy properties, ranging from the evolution of the SMF to quantifying the impact of positive AGN feedback across galaxies with varying statuses (central or satellite) and stellar masses. In the subsequent sections, we detail the predictions of both the ModA and ModB models for each property, clarifying whether these are direct predictions or outcomes derived from the calibration process (evolution of the SMF from redshift $z=3$ to the present epoch).

\begin{figure}
\begin{center}
\begin{tabular}{cc}
\includegraphics[width=0.41\textwidth]{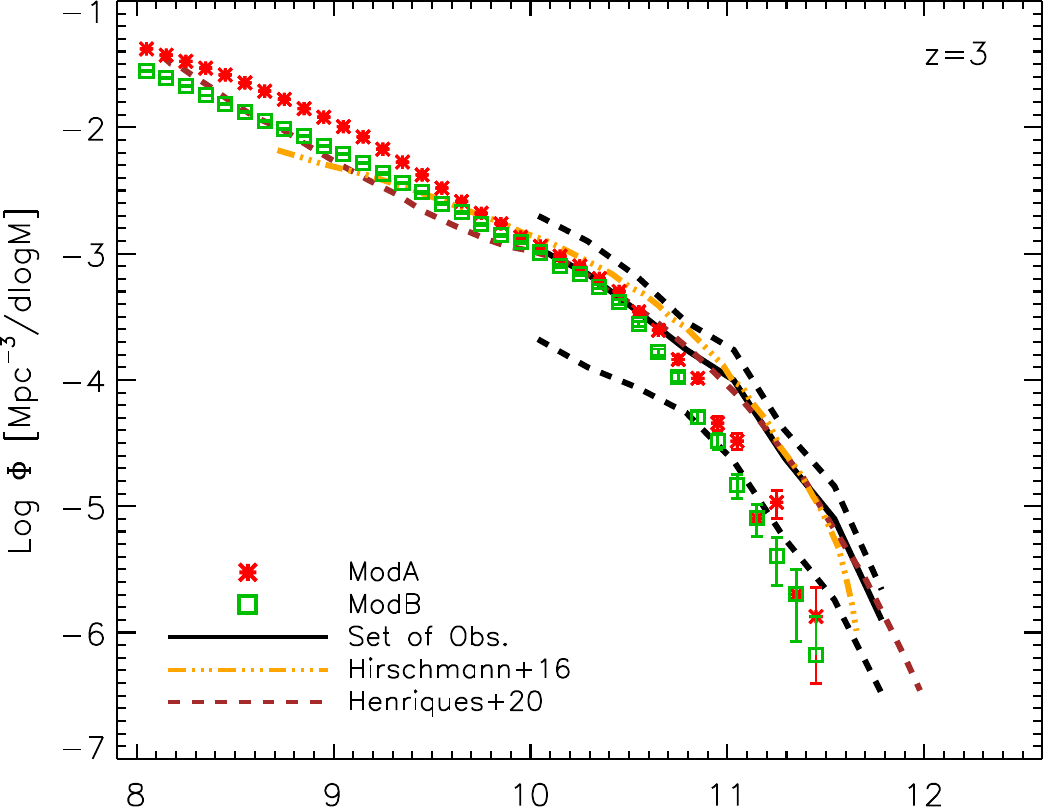} \\
\includegraphics[width=0.41\textwidth]{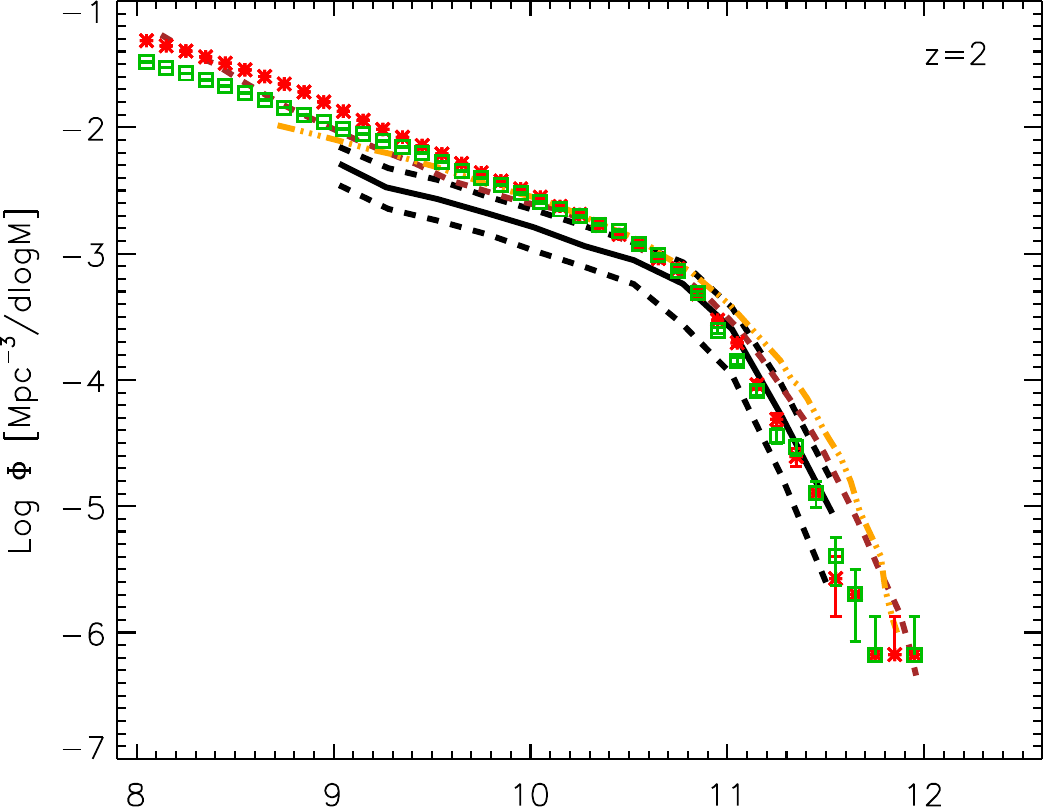} \\
\includegraphics[width=0.41\textwidth]{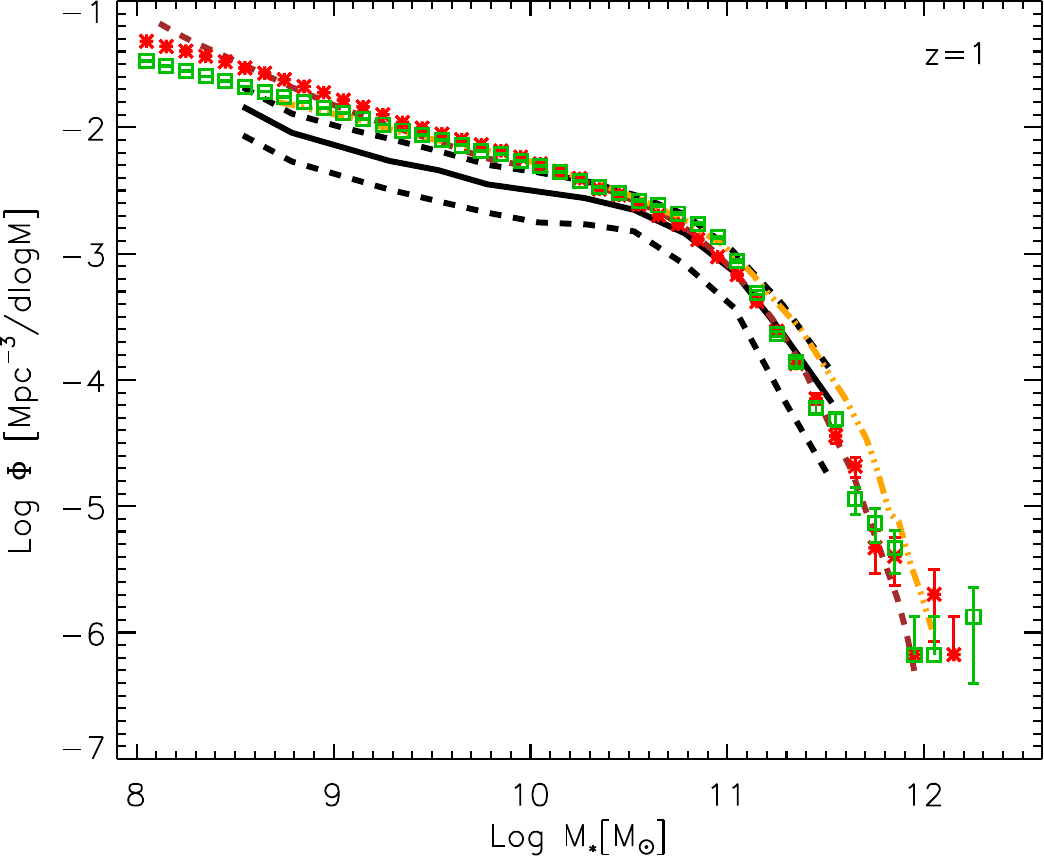} \\
\end{tabular}
\caption{Evolution of the SMF from $z=3$ to $z=1$, as predicted by ModA (represented by red stars) and ModB (represented by green squares), compared with the set of observations upon which FEGA has been calibrated (illustrated by black lines) and described in Section \ref{sec:set_obs}. Additionally, we compare our predictions with those of two SAMs that feature the same SN feedback as FEGA: the model by \citealt{hirschmann2016} (depicted by an orange line), which aligns closely with ModA, and the model by \citealt{henriques2020} (depicted by a brown line), which aligns closely with ModB. Both ModA and ModB reasonably describe the evolution of the SMF down to $z=1$. The effectiveness of the stellar feedback in ModB is evident even at high redshifts. Although ModA exhibits redshift-dependent stellar feedback (stronger at higher redshifts), the two models show differences for stellar masses below $\log M_* \sim 9$.}
\label{fig:SMF_hz}
\end{center}
\end{figure}

\begin{figure*}
\centering
\includegraphics[width=0.8\textwidth]{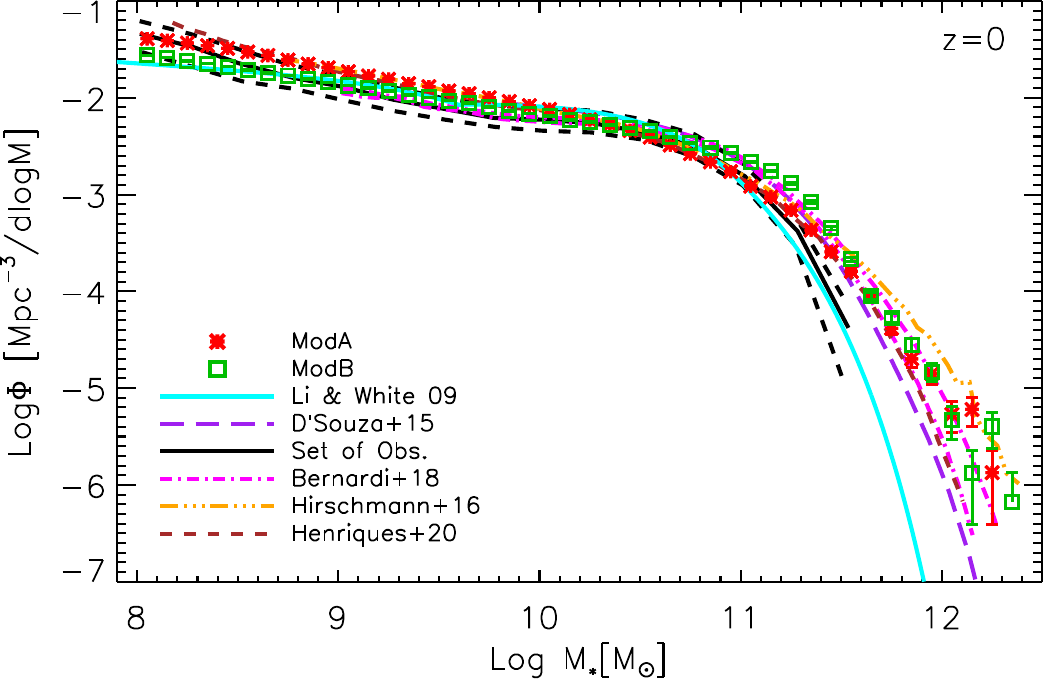}
\caption{SMF at the present time, as predicted by ModA and ModB, depicted using the same styles and colors as shown in Figure \ref{fig:SMF_hz}. We compare the predictions of these two models with the set of observations used for calibration, as well as with the predictions of \citealt{hirschmann2016} and \citealt{henriques2020} (represented by the same colors as in the previous figure). Additionally, we consider another set of observations from different authors: \citealt{li-white2009} (represented by a cyan line), \citealt{dsouza2015} (represented by a purple line), and \citealt{bernardi2018} (represented by magenta lines indicating the $\pm\sigma$ region). It should be noted (see the text for further details) that at $z=0$, the calibration encompasses the region of the SMFs by \citealt{bernardi2018} for stellar masses $\log M_* \gtrsim 11$. Both ModA and ModB predict the observed SMF accurately, with both models favoring the high-mass end of \citealt{bernardi2018}'s SMF. This has important consequences for the stellar-to-halo mass relation.}
\label{fig:SMF_z0}
\end{figure*}

\subsection{Stellar Mass Function}\label{sec:SMF}
The evolution of the SMF is illustrated in Figures \ref{fig:SMF_hz} and \ref{fig:SMF_z0}. It is important to note that {\small FEGA}, in both ModA and ModB versions, is designed to offer a reliable representation of the SMF's evolution, given that it was calibrated based on this metric. Throughout this paper, red/green symbols and lines correspond to ModA/ModB (unless otherwise specified). Conversely, symbols or lines in colors other than red and green pertain to either observed data or predictions from alternative models.

Figure \ref{fig:SMF_hz} showcases the SMF at redshifts $z=3$, $z=2$, and $z=1$, juxtaposing ModA and ModB with the observation set utilized during calibration, the model zDEP from \citealt{hirschmann2016} (referred to as H16) featuring identical stellar feedback, and the model from \citealt{henriques2020} (referred to as H20). Overall, all models demonstrate a relatively good fit to the SMF evolution. The primary discrepancy between ModA/B and H16 or H20 is evident at the high-mass end of the $z=3$ SMF. Specifically, both ModA and ModB appear to underestimate this segment compared to the others. The most likely explanation for it comes from the size of our simulation, which might not be sufficiently large to capture a representative sample of massive galaxies at high redshifts. Moreover, it is crucial to emphasize that our data represent the raw number densities generated by the models, not fitted values.

Figure \ref{fig:SMF_z0} presents the SMF at the current epoch, with our models compared against several observational datasets, H16, and H20 models. The figure displays the calibration-used observations (black lines), the SMF by \cite{li-white2009} (cyan line), the SMF by \cite{dsouza2015} (purple line), and the SMF by \cite{bernardi2018} (magenta lines). While the observations align well from the low-mass end up to the knee at $\log M_* \sim 11$, discrepancies emerge towards higher masses, leading to markedly divergent number densities for the most massive galaxies. This divergence is a critical point we revisit in Sections \ref{sec:sthmass} and \ref{sec:overview}. Notably, all models—including ours, H16, and H20—align better with the high-mass end from \cite{bernardi2018}.

Setting this aside, both ModA and ModB offer a satisfactory depiction of the SMF's evolution. We proceed to investigate their capability to capture properties for which they were not explicitly calibrated.

\begin{figure*}
\centering
\includegraphics[width=0.95\textwidth]{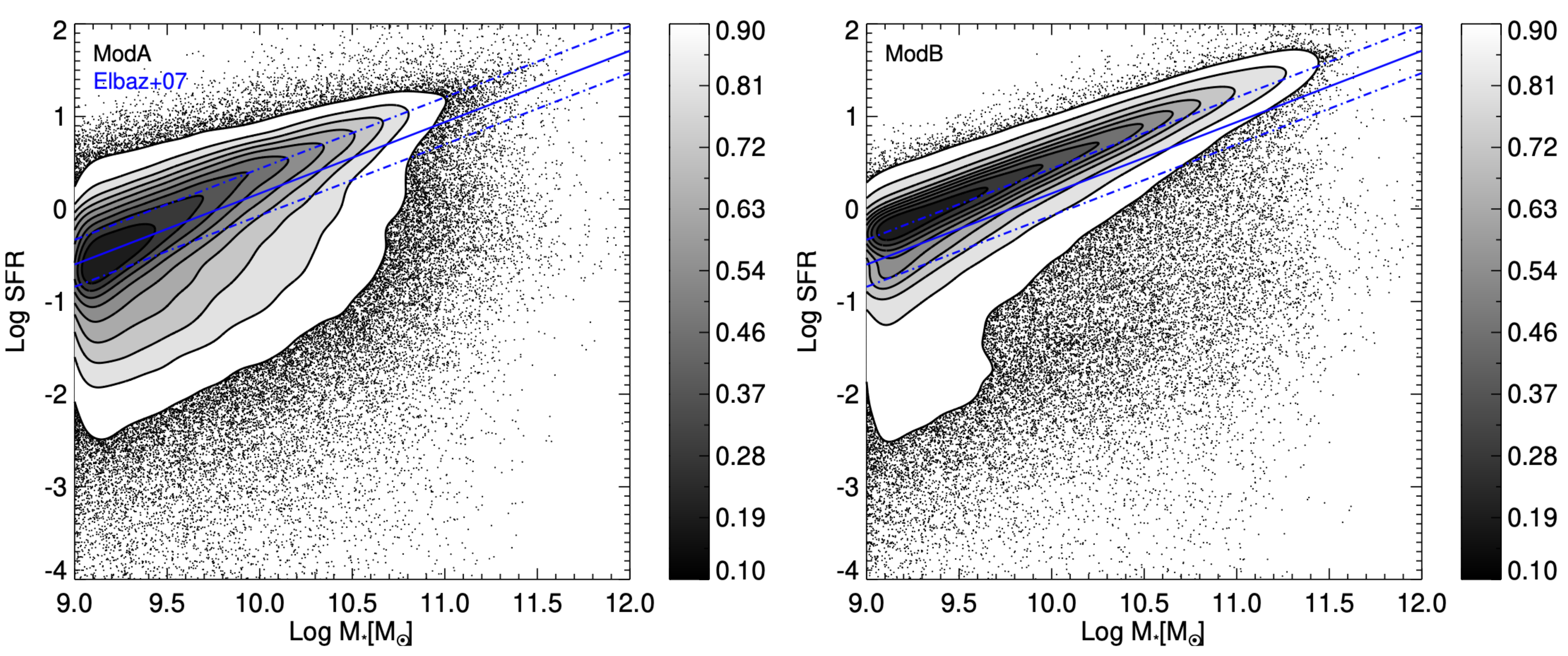}
\caption{The star formation rate-stellar mass relation at $z=0$ predicted by ModA is shown in the left panel, while that predicted by ModB is shown in the right panel. These are compared with the observed main sequence data by \citealt{elbaz2007} (represented by blue lines). The contours delineate different percentages of the data, as indicated by the colors in the accompanying bar. The last contour encloses 90\% of the full sample, with the remaining 10\% represented by individual black dots. Overall, ModA aligns well with the observed main sequence, whereas ModB appears to be slightly biased towards higher values but still falls within the observed scatter. Notably, ModA predicts a broader red sequence that extends towards massive galaxies, while the red sequence predicted by ModB is more concentrated towards less massive galaxies.}
\label{fig:sfr_mass}
\end{figure*}

\begin{figure}
\centering
\includegraphics[width=0.45\textwidth]{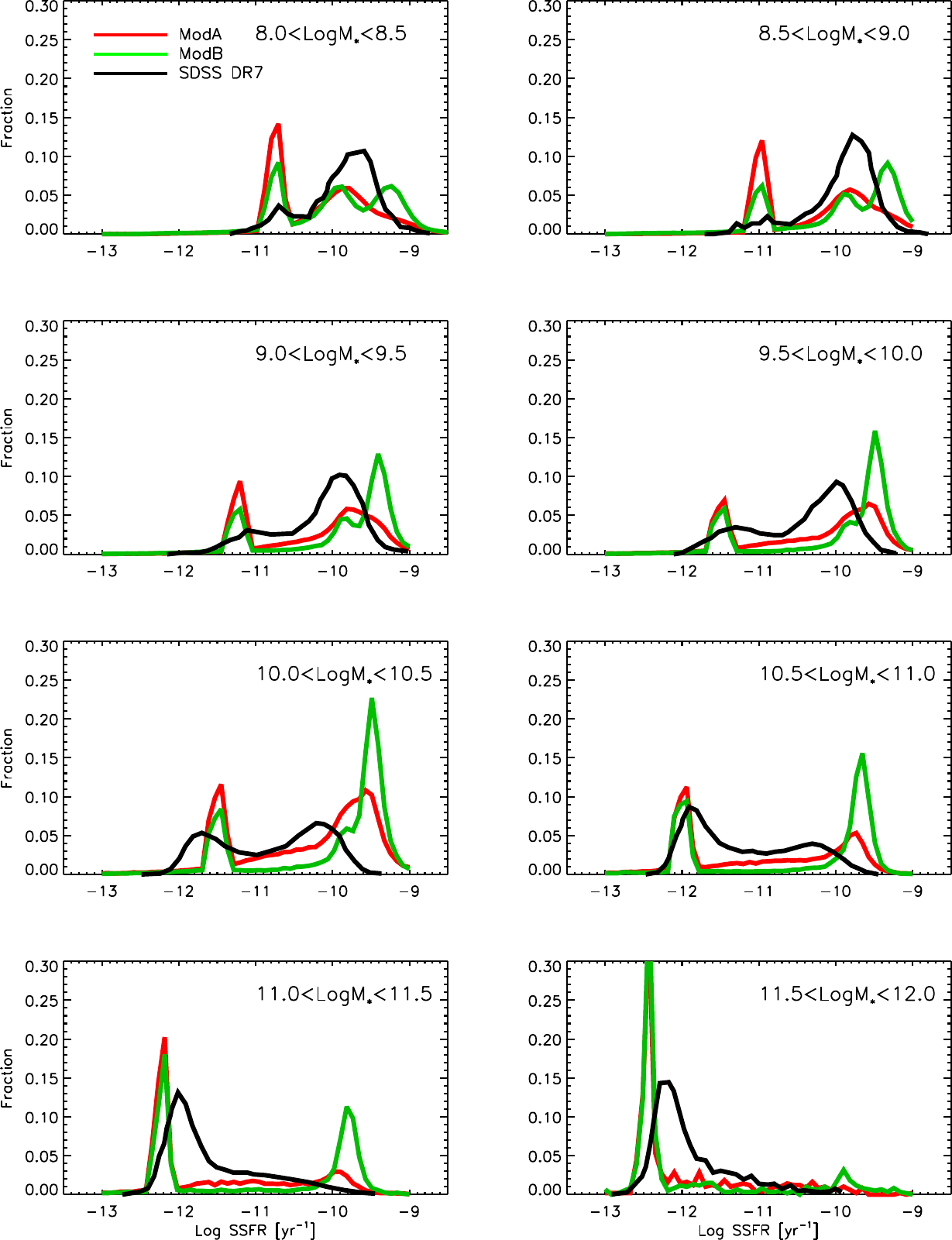}
\caption{Specific star formation rate distributions across different ranges of stellar mass are displayed in various panels, as indicated in the legend, for both ModA and ModB. These distributions are compared with the equivalent distribution derived from SDSS DR7, represented by black lines. The two models predict distinct distributions, particularly in the range of $10.5 < \log M_* < 11.0$, where ModA appears to align more closely with the observed distributions. Overall, there are no drastic differences observed across either range of stellar mass.}
\label{fig:ssfr_distr}
\end{figure}

\subsection{Star and Specific Star Formation Rates}\label{sec:SFRmass}
Star formation is a cornerstone in galaxy formation, and a SAM should ideally yield star formation rates closely mirroring the observed rates. Figure \ref{fig:sfr_mass} portrays the SFR-stellar mass relation as predicted by ModA (left panel) and ModB (right panel) at $z=0$, placed together with the observed main sequence by \citealt{elbaz2007} (represented by blue lines). Each shell between two successive contours in the panels encapsulates 10\% of the data, culminating in the last contour, which encompasses 90\% of the data. The residual 10\% are represented by individual black dots.

Both models generally capture the observed main sequence, yet ModA exhibits slightly superior performance compared to ModB. Specifically, while ModA's main sequence aligns with the observed one ($1\sigma$ scatter included), ModB leans slightly above the scatter depicted by the dashed line. However, ModB still remains within the observed $2\sigma$ scatter. A notable divergence between the two models emerges in the red sequence, or the domain occupied by red galaxies. ModA forecasts a broader red sequence than ModB, which encompasses more massive galaxies. Additionally, by examining the interval between successive contours, galaxies in ModB appear to be bluer than those in ModA. The implications of this divergence require further scrutiny, particularly after placing together the SSFR distribution in stellar mass bins with observed data.

In Figure \ref{fig:ssfr_distr}, we examine this aspect by segmenting our galaxy sample into eight stellar mass ranges (different panels), ranging from $8.0<\log M_* <8.5$ to $11.5<\log M_* <12.0$ (we do not show the SSFR distribution of more massive galaxies because of the lack of the observed counterpart in that range of stellar mass). The predictions of our models are contrasted with the observed distribution from SDSS-DR7 (depicted by black lines) within each range. Notably, both models generally pinpoint the peaks of each distribution, exhibiting varying degrees of accuracy across ranges. However, ModB consistently overshadows ModA in the height of peaks for star-forming and blue galaxies. Conversely, with exceptions in the first two ranges (lowest mass bins), the peak heights for red galaxies closely resemble each other across both models. This suggests that galaxies in ModB lean towards bluer colors compared to ModA, and also relative to observed galaxies. It is essential to emphasize that neither the SFR-mass relation nor the SSFR distribution were included in the calibration process.

\begin{figure*}
\begin{center}
\begin{tabular}{cc}
\includegraphics[width=0.46\textwidth]{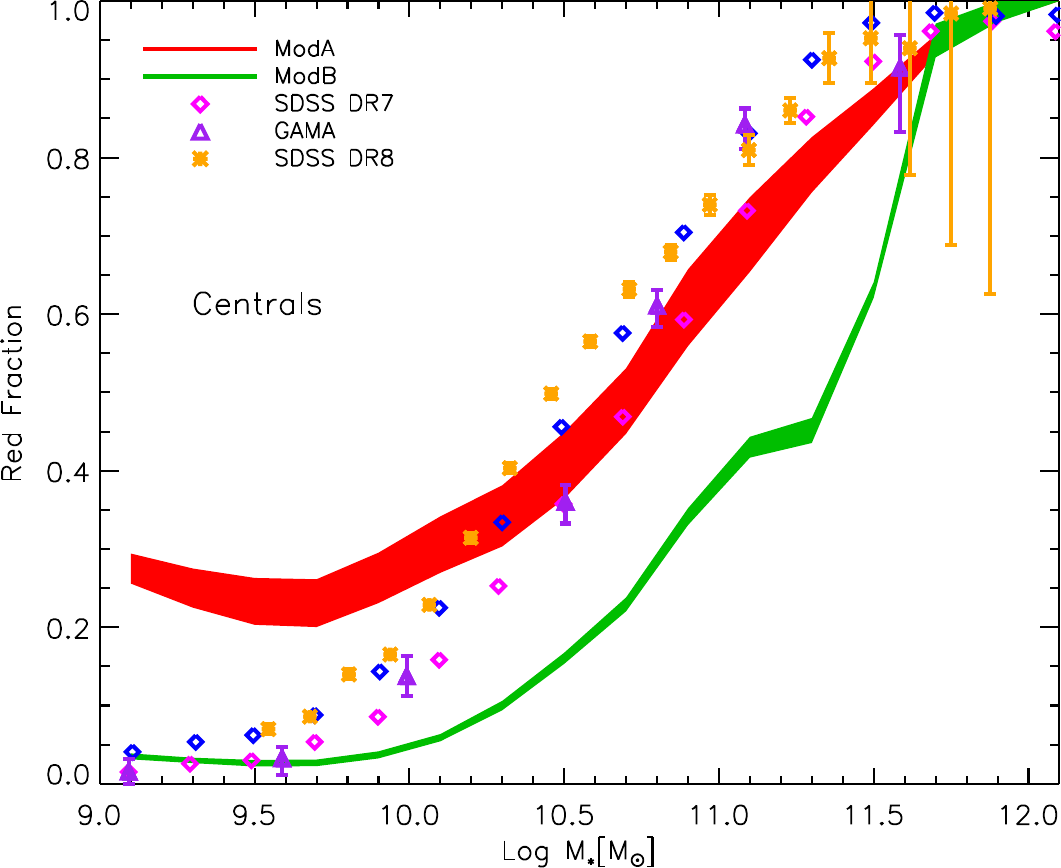} &
\includegraphics[width=0.46\textwidth]{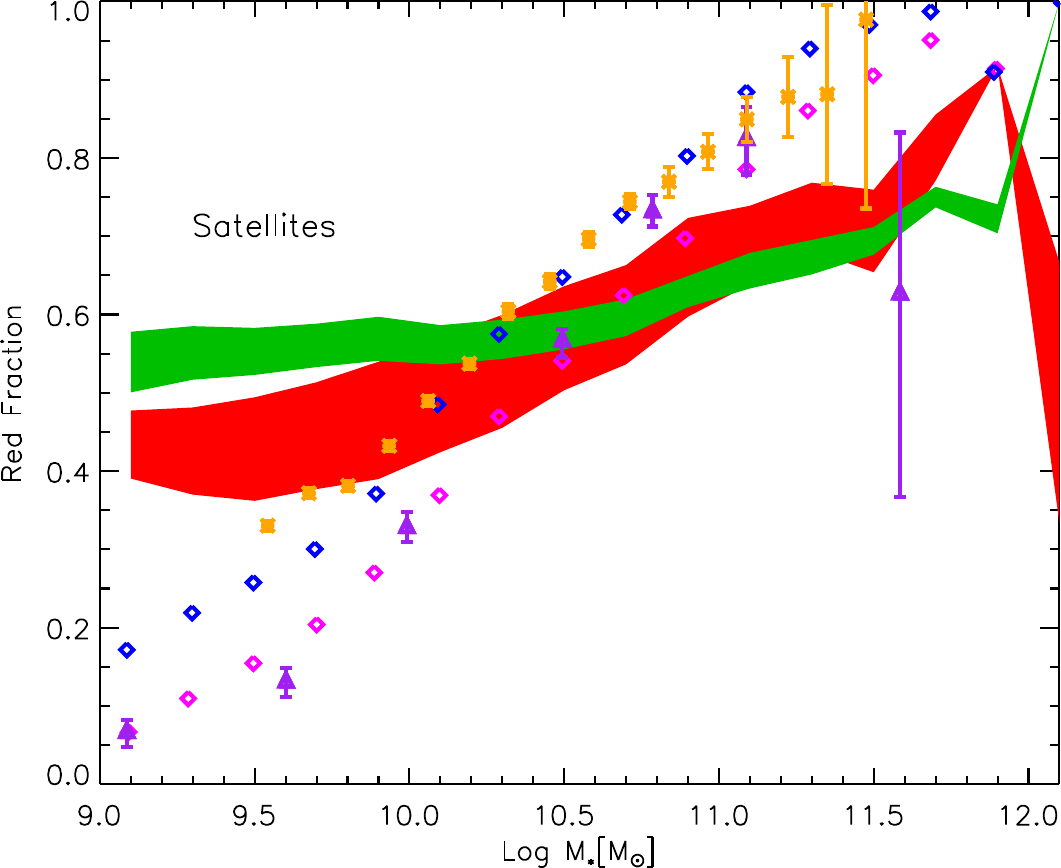} \\
\end{tabular}
\caption{The fraction of red galaxies as a function of stellar mass is displayed for centrals in the left panel and for satellites in the right panel. Our model predictions from ModA (represented by the red region) and ModB (represented by the green region) are compared with observed fractions from SDSS-DR7 (\citealt{delucia2024}, represented by magenta and blue diamonds), SDSS-DR8 (\citealt{hirschmann2014}, represented by orange stars), and from the GAMA survey by \citealt{davies2019} (represented by purple triangles). Red galaxies are selected based on different cuts in $\log$SSFR, specifically: -10.5 for GAMA data, and -10.6 and -11.0 for SDSS data. The regions represented by red and green in our models encompass the smallest and largest SSFR cuts, respectively. The plots highlight an important distinction between the two models. While ModA aligns more closely with the observed data in both cases, both models capture the general trend of increasing fraction with increasing mass. However, their predictions are higher than expected for low-mass galaxies and lower for high-mass galaxies. This discrepancy is more pronounced for satellite galaxies, particularly for ModB.}
\label{fig:red_fraction}
\end{center}
\end{figure*}

\begin{figure}
\centering
\includegraphics[width=0.45\textwidth]{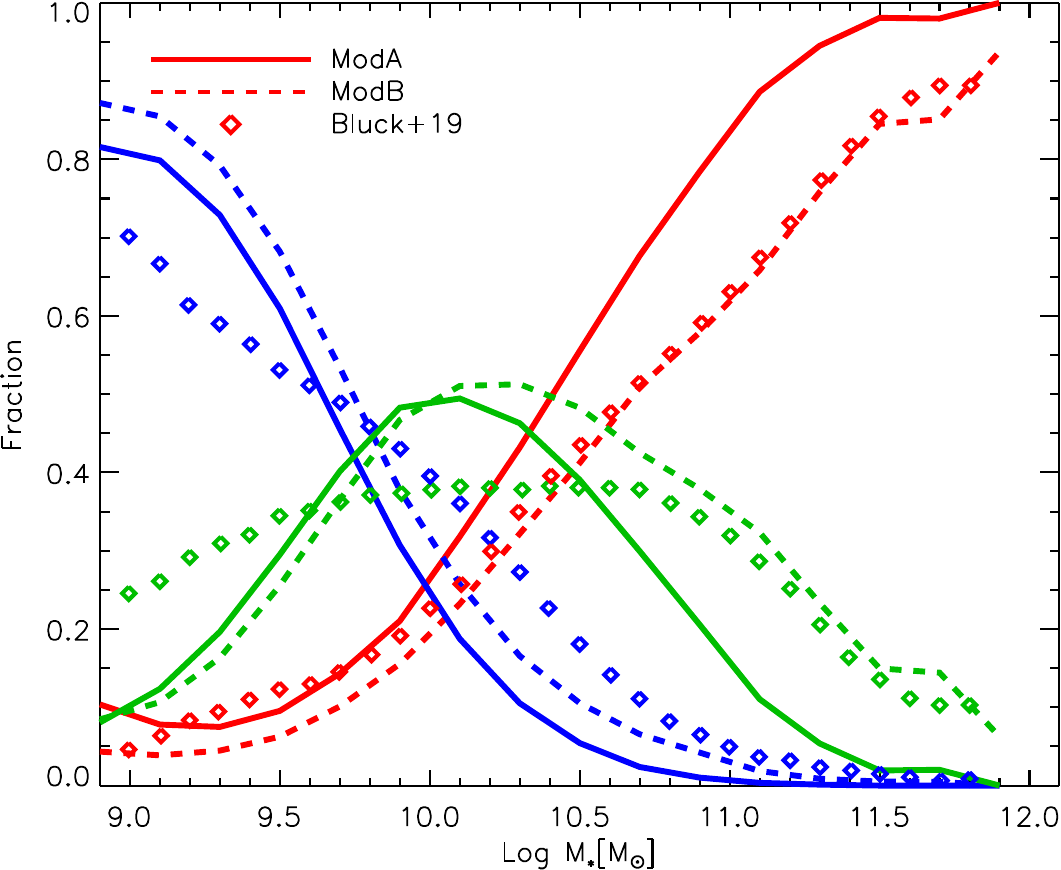}
\caption{Fraction of galaxies with different morphologies, classified based on the bulge-to-total ratio (B/T), plotted as a function of stellar mass. Galaxies are categorized as bulge-dominated if they have $\rm{B/T} > 0.7$ (represented by red), disk-dominated if $\rm{B/T} < 0.3$ (represented by blue), and spheroidals for values in between (represented by green). Our predictions from ModA are shown as solid lines, while those from ModB are represented by dashed lines. These predictions are compared with observed data from \citealt{bluck2019}, represented by diamonds. Remarkably, both models demonstrate a high level of accuracy in reproducing the observed trends, which is unprecedented in SAMs.}
\label{fig:morphology}
\end{figure}

\subsection{Fraction of Red Galaxies and Morphology}\label{sec:redfrac}
The disparity in colors between galaxies in ModB and ModA, with ModB leaning bluer than both ModA and observed galaxies, prompts an explicit comparison using the fraction of red galaxies as a function of stellar mass, further differentiated between centrals and satellites. This comparison is presented in Figure \ref{fig:red_fraction}, with the left panel focusing on centrals and the right panel on satellites. Our models are compared with observed fractions from SDSS-DR7 (\citealt{delucia2024}, magenta and blue diamonds), SDSS-DR8 (\citealt{hirschmann2014}, orange stars), and from the GAMA survey by \cite{davies2019} (purple triangles). To align with observations, red galaxies are identified based on $\log$SSFR cuts, with the red and green regions encapsulating the three utilized cuts: -10.5 yr$^{-1}$ for GAMA data, and -10.6 yr$^{-1}$ and -11.0 yr$^{-1}$ for SDSS data.

Both ModA and ModB capture the observed upward trend in the fraction of red galaxies with increasing stellar mass for both centrals and satellites. However, ModA aligns more closely with observed fractions compared to ModB, with the divergence between the models becoming more pronounced for central galaxies. Despite this, both models tend to overestimate the fraction in low-mass galaxies and underestimate it in larger galaxies. Exceptionally, the model by \cite{delucia2024} aligns almost perfectly with both observed fractions. {\small FEGA} can replicate these fractions with comparable or even superior accuracy relative to other SAMs. The reasons underlying the discrepancies between the models and observations will be explored in Section \ref{sec:overview}.

Another historical shortcoming of SAMs pertains to galaxy morphology. Setting aside the intricacies in defining \emph{galactic morphology}, only a handful of SAMs (e.g., \citealt{irodotou2019,izquierdo-villalba2019}) have managed to reasonably replicate the fraction of galaxies with varying morphologies based on their bulge-to-total ratio (B/T), a commonly adopted proxy for morphology. Figure \ref{fig:morphology} addresses this, illustrating the fractions of bulge-dominated galaxies ($\rm{B/T}>0.7$, in red), disk-dominated galaxies ($\rm{B/T}<0.3$, in blue), and spheroidals (those in between, in green) for ModA (solid lines) and ModB (dashed lines). Our predictions are compared with observational data by \cite{bluck2019}, with colors mirroring the B/T ranges used in our models. Our models demonstrate an unprecedented accuracy in reproducing observed trends and fractions across wide ranges. Notably, ModB is more in line with observations than ModA.

The enhanced fidelity of our SAM in replicating galaxy morphologies stems from our novel treatment of disk instability. Specifically, the critical mass in Equation \ref{eqn:di_mcrit} is reduced by a factor of 5 in both models (refer to Table \ref{tab:parameters} for the $\delta_{\rm{DI}}$ values), a modification aligning more closely with recent model parameters (e.g., \citealt{lagos2018,izquierdo-villalba2019}). It is pivotal to reiterate that neither this relation nor the fraction of red galaxies were included in the calibration process.

\begin{figure}
\centering
\includegraphics[width=0.45\textwidth]{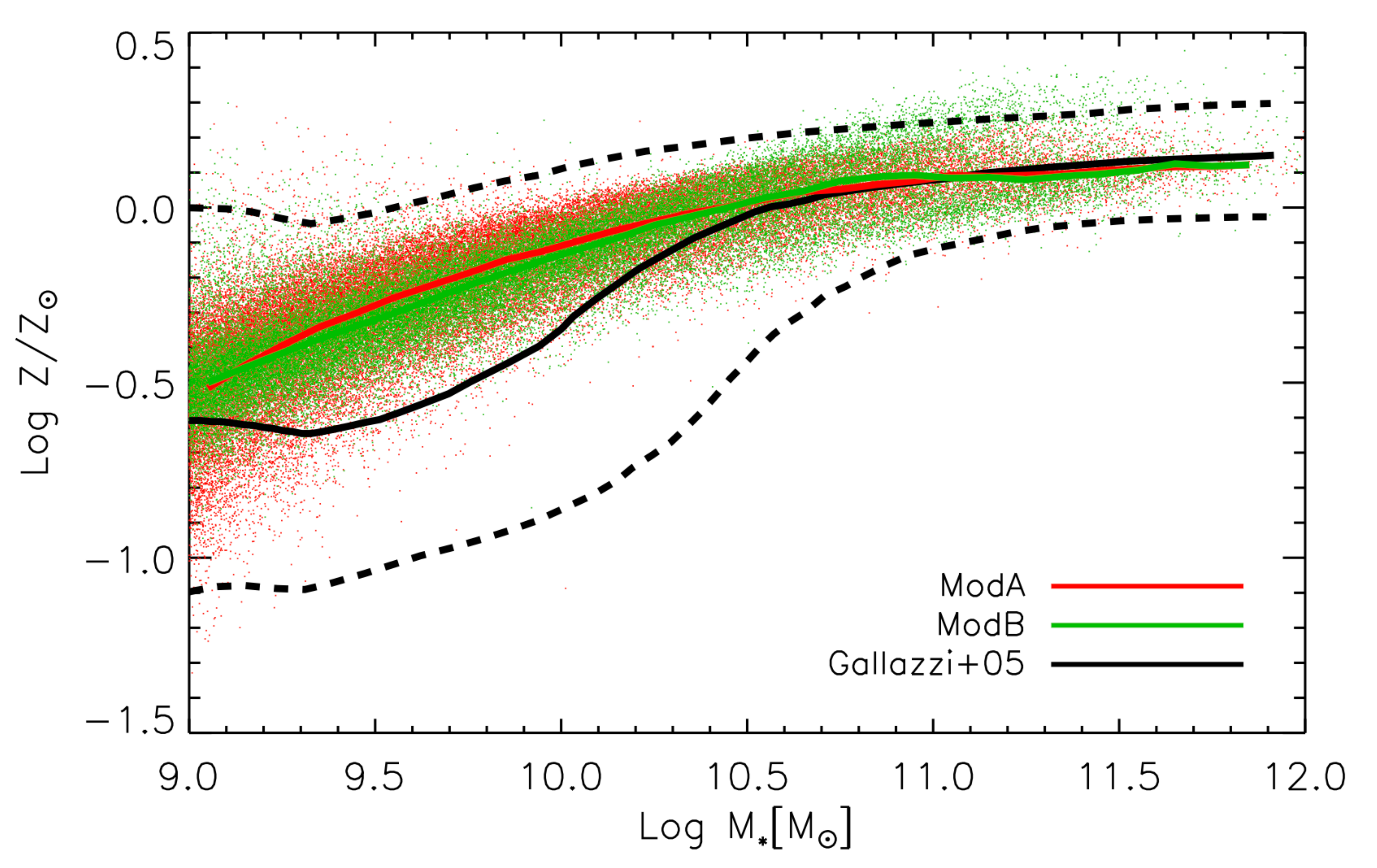}
\caption{The metallicity-stellar mass relation predicted by ModA is represented by red lines and dots, while that predicted by ModB is shown with green lines and dots. These predictions are compared with the observed relation from \citealt{gallazzi2005}, depicted by black lines. While both models capture the general trend of increasing metallicity with stellar mass, neither accurately reproduces the local slopes at masses lower than $\log M_* \sim 10$. Nevertheless, the majority of data points from both models fall within the observed scatter.}
\label{fig:mm_relation}
\end{figure}

\subsection{Mass-Metallicity Relation}\label{sec:metallicity}
The mass-metallicity relation is a focal aspect rigorously constrained by SAMs, and {\small FEGA} is expected to perform similarly. The metallicity of stars hinges on several processes: primarily star formation, which dictates the fraction of metals produced (refer to Table \ref{tab:parameters}), SN feedback that can expel metals from low-mass galaxies (as discussed in H16), the stripping of stars during the formation of the ICL (as discussed in \citealt{contini2019}), among others.

Figure \ref{fig:mm_relation} depicts the average mass-metallicity relation projected by our models (indicated by red and green lines), individual data points (in red and green), and the observed relation (in black lines) as per \cite{gallazzi2005}. The fraction of metals produced per star has been adjusted post-calibration for both models to align with the average observed metallicity in massive galaxies. As discerned from Table \ref{tab:parameters}, ModA necessitates half the fraction demanded by ModB, specifically 0.025 compared to 0.05, and their mean relations coincide. Both models can emulate the observed relation, exhibiting no substantial differentiation between them, except for ModB displaying slightly reduced scatter in data points.

Nevertheless, the models fall short in capturing the curvature evident in the observed relation towards low-mass galaxies, albeit they both remain within the observed scatter. It is crucial to acknowledge that this discrepancy might stem from the specific observation chosen for comparison. In fact, as elucidated in H20, the SDSS-DR7 data from \cite{zahid2017} do not exhibit the same trend. A noteworthy caveat is that in the initial version of {\small FEGA}, i.e., prior to calibration, $f_{\rm{metals}}$ was fixed at 0.03, a prevalent value in use. To avoid introducing an additional parameter for calibration, we opted to adjust it post-calibration as explained earlier. While this value remains acceptable for ModA, it is inadequate for ModB, causing the model to markedly underestimate the metallicity of massive galaxies. Indeed, in the calibration of the model by \cite{henriques2013}, which also incorporates robust SN feedback akin to our ModB, $f_{\rm{metals}}$ was determined to be 0.047.

\begin{figure*}
\centering
\includegraphics[width=0.95\textwidth]{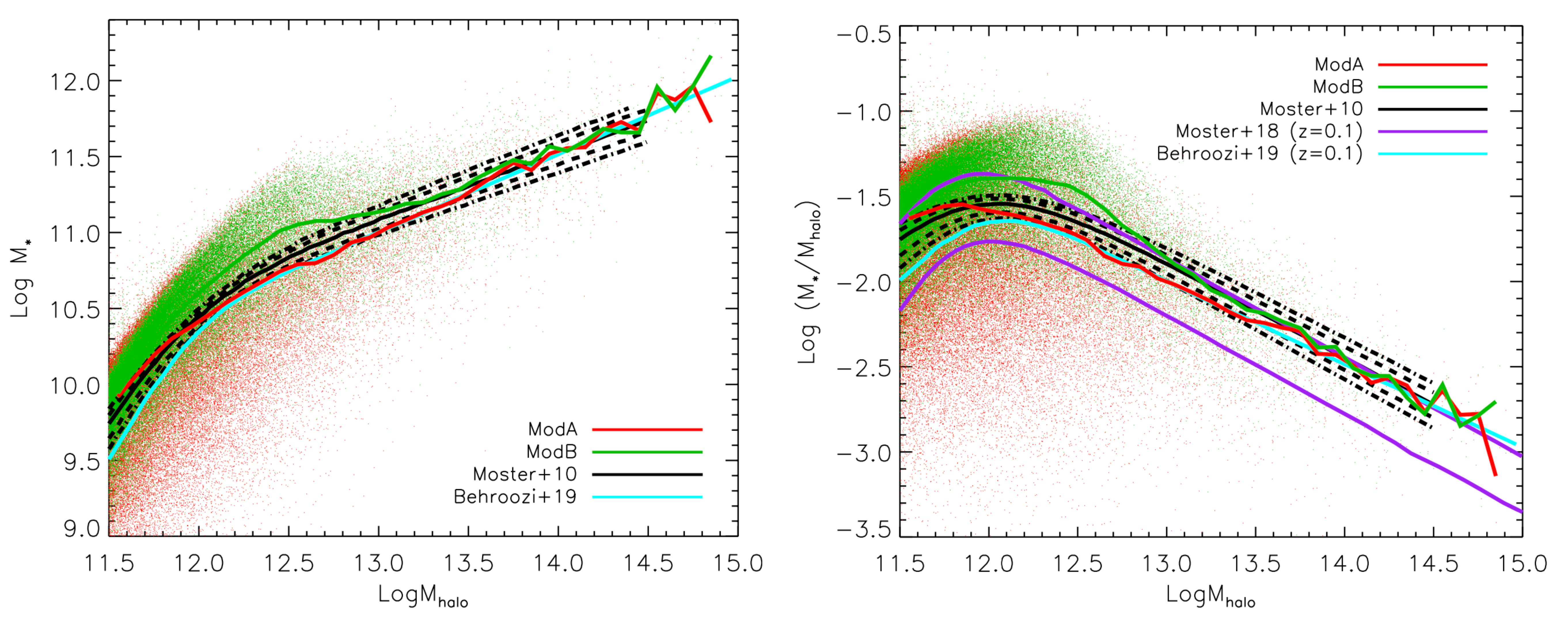}
\caption{In the left panel, we compare the stellar mass versus halo mass relation predicted by our models, ModA and ModB, with those from the models of \citealt{moster2010} (represented by black lines), \citealt{moster2018} (represented by purple lines), and \citealt{behroozi2019} (represented by cyan lines). The right panel compares the stellar-to-halo mass ratio versus halo mass. An apparent difference between ModA and ModB is evident in these plots. While ModA closely aligns with the relation predicted by Moster+10, ModB diverges, showing an increase relative to ModA in low-mass halos ($\log M_{\text{halo}} < 12.7$). This divergence exceeds the $2\sigma$ scatter represented by the dash-dotted black lines. Additionally, ModB exhibits a narrower scatter, particularly in low-mass halos. A key takeaway from this figure is that to reproduce the predicted relation on cluster scales, the high-mass end of the SMF at $z=0$ must closely match that observed by \cite{bernardi2018}. Further discussion on this topic can be found in Section \ref{sec:overview}.}
\label{fig:mstar_mhalo}
\end{figure*}

\subsection{Stellar-Halo Mass Relation}\label{sec:sthmass}
The stellar budget within dark matter halos is rigorously characterized by theoretical models, as evidenced by various studies such as \cite{moster2010}, \cite{behroozi2013}, \cite{moster2018}, \cite{behroozi2019}, among others. The normalized stellar-to-halo mass relation, which essentially portrays the efficiency of star formation across halos of varying masses, is notably observed to peak around Milky Way-like halos, as referenced in the aforementioned works. These relations hinge on a myriad of physical processes that either augment or diminish the stellar mass within galaxies, encompassing star formation, mergers, tidal stripping, and so forth. Ideally, achieving the alignment of galaxies with their respective dark matter halos would mark a significant triumph for a SAM.

Figure \ref{fig:mstar_mhalo} delineates the stellar-to-halo mass relation on the left panel and its halo mass normalized counterpart on the right, as forecasted by our models, ModA and ModB, juxtaposed with various theoretical predictions as annotated in the legend. While both models do capture the anticipated relations, distinct discrepancies emerge between them. Primarily, ModB exhibits a tighter scatter, especially in low-mass halos, compared to ModA. Secondly, across both relations, ModB portrays an elevated average stellar mass and stellar-to-halo mass ratio for low-mass scales ($\log M_{\rm{halo}} \leq 12.7$), while ModA leans within the scatter of these predictions.

A noteworthy consistency between both models is evident at higher halo mass scales. Remarkably, both models precisely mirror the expected relations, aligning with the average projected values at $\log M_{\rm{halo}} \geq 13.0$. However, it is pivotal to recognize that ModA's predictions align with the projected values across the entire range of halo masses explored. Another salient observation, which will be discussed comprehensively in Section \ref{sec:overview}, pertains to the implications of this precise alignment between the models and analytical predictions. This high level of congruence in matching the high halo mass end of these relations suggests that the favored high mass end of the SMF at present times ought to correspond to the range observed by \cite{bernardi2018}. Deviating from this would either result in underestimating or overestimating the theoretical predictions for these relations at larger halo masses.

\begin{figure*}
\centering
\includegraphics[width=0.95\textwidth]{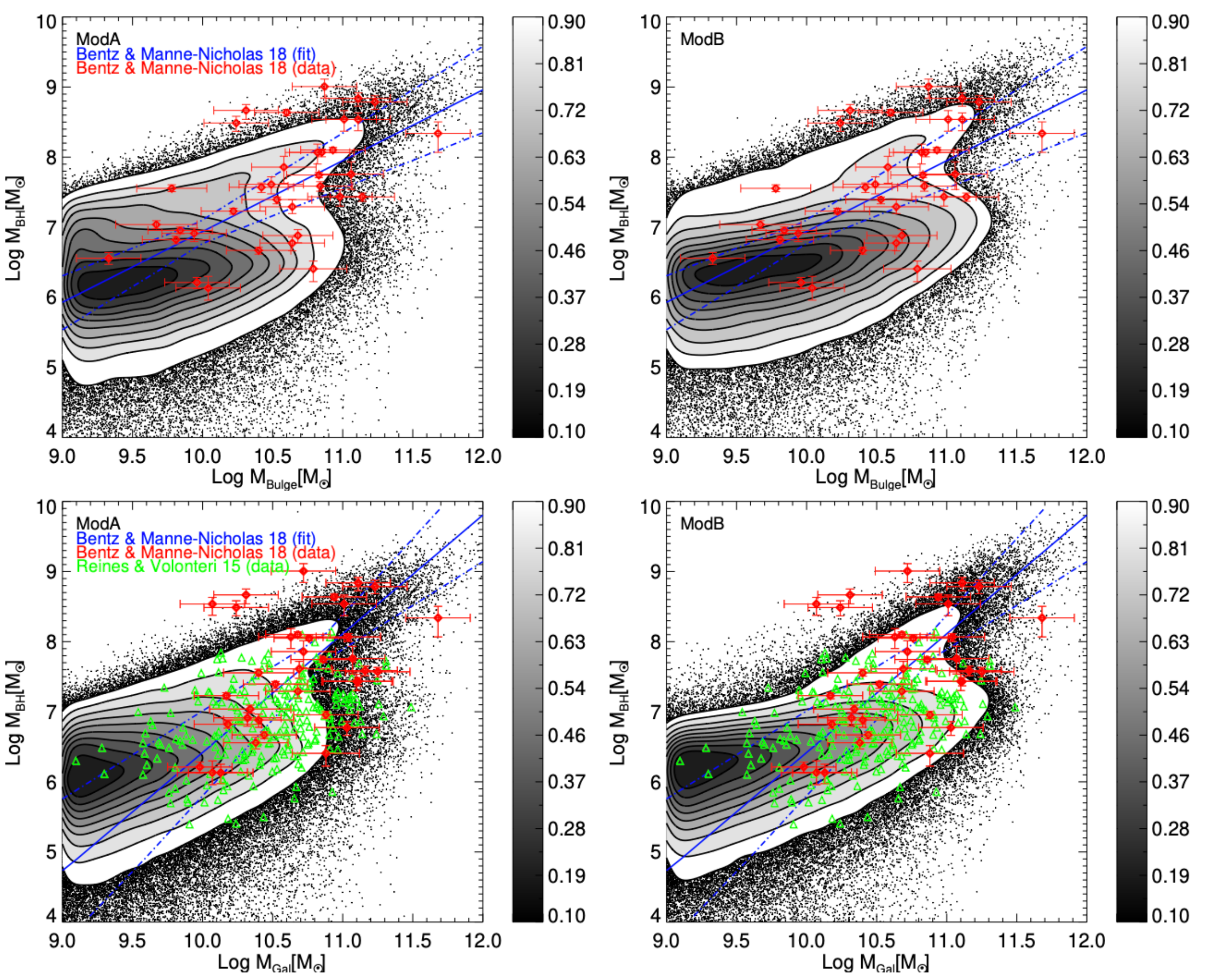}
\caption{In the top panels, we compare the BH versus bulge mass relations predicted by ModA (left) and ModB (right) with observed data and relations from \citealt{bentz-manne2018}, represented by blue lines and red diamonds. In the bottom panels, the BH versus stellar mass relations are compared with observed data from \citealt{reines2015}, depicted by green triangles. Contours in the plots indicate the percentages of data points they enclose, color-coded as shown in the accompanying bar up to 90\%. Data points not included within the contours are represented by individual black dots, accounting for the remaining 10\%. Both models are fairly capable of describing both relations in a similar manner. However, the second relation is over-predicted at low stellar masses compared to the observations. It should be noted that these observed relations are fitted to data points, the majority of which lie above the low-mass range.}
\label{fig:bh_mass}
\end{figure*}

\begin{figure}
\centering
\includegraphics[width=0.9\textwidth,angle=270]{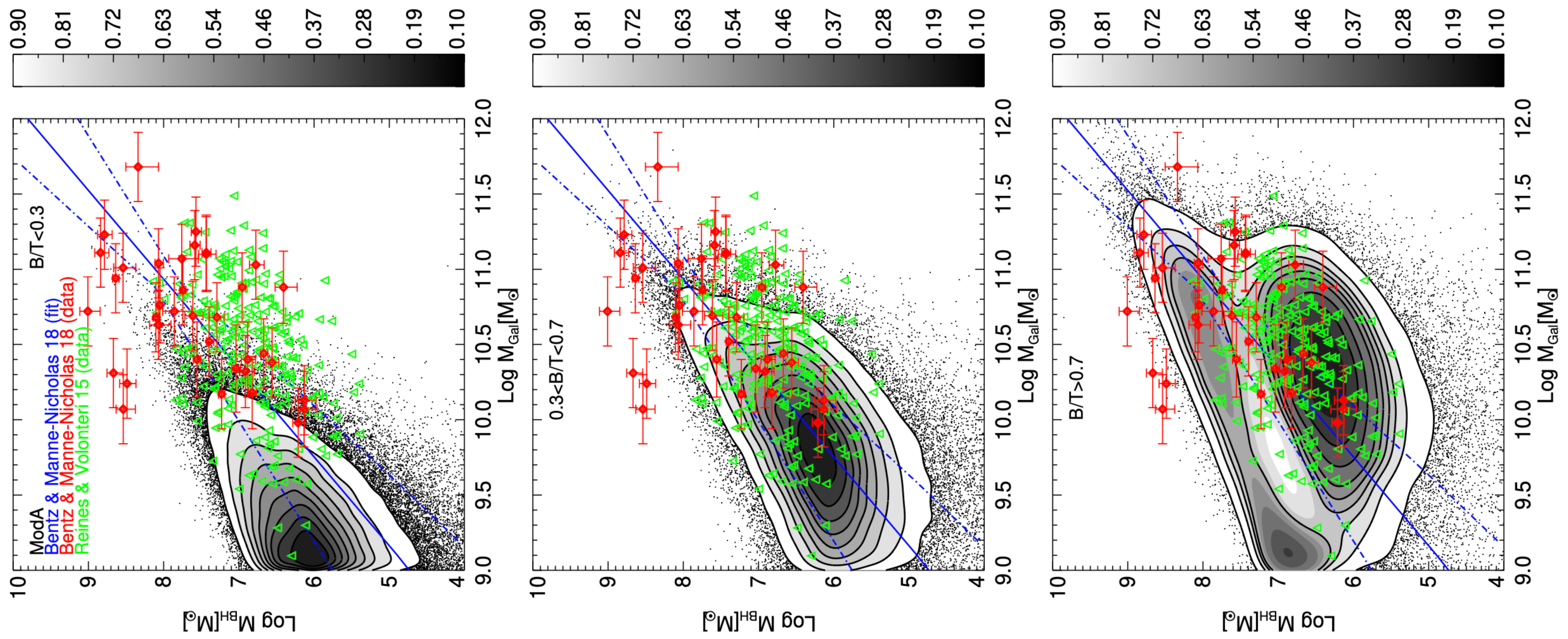}
\caption{BH versus stellar mass relations predicted by ModA (with similar trends shown by ModB) presented for galaxies of different morphologies in three panels. The top panel displays the relation for disk galaxies with $B/T < 0.3$, the middle panel shows the relation for lenticulars with $0.3 < B/T < 0.7$, and the bottom panel depicts the relation for ellipticals with $B/T > 0.7$. These model predictions are compared with the same observed data as shown in the bottom panels of Figure \ref{fig:bh_mass}. The plots reveal distinct behaviors for each morphology type. Disk galaxies appear slightly biased high compared to observations, following a similar relation but with a higher intercept. Lenticular galaxies closely follow the observed relation. Elliptical galaxies, however, exhibit two distinct populations: one that adheres to a tight correlation different from the observed one, and another that aligns with the observed data cloud from \cite{reines2015}. This underscores the importance of galaxy morphology in understanding the black hole versus stellar mass relation.}
\label{fig:bh_mass2}
\end{figure}

\subsection{Black Hole-Bulge/Galaxy Mass Relations}\label{sec:bhmass}

The relationship between the BH mass and the bulge or stellar mass of a galaxy is a crucial indicator of the BH growth. Numerous studies in the past, including works by \cite{magorrian1998,haring2004,ferrarese2005,graham2007,mcconnell-ma2013,kormendy-ho2013,reines2015,bentz-manne2018,davies2018,baron-menard2019}, have delved into these relationships. The consensus from these studies suggests a strong correlation between the BH mass and either the bulge or stellar mass of a galaxy, indicating an intrinsic evolutionary link between the formation of galactic bulges and the growth of BHs.

In this section, we scrutinize the predictions of our SAM, emphasizing its capability to accurately represent these relationships when placed together with observed data. However, it is worth noting that not all galaxies adhere to the same relation, as highlighted by \cite{reines2015}. Therefore, a potential distinction based on galaxy morphologies (\citealt{davies2018}) could be pivotal, and we intend to explore this aspect.

Figure \ref{fig:bh_mass} presents the BH versus bulge mass (top panels) and stellar mass (bottom panels) relationships as projected by ModA (left panels) and ModB (right panels). These predictions are compared against observed data and relations by \cite{bentz-manne2018} (blue lines and red diamonds) and the data from \cite{reines2015} (green triangles). Similar to previous figures, the contours delineate the distribution percentages of the data points, color-coded as per the bar, reaching up to 90\%, while the remaining 10\% is denoted by individual black dots.

Both models aptly reproduce these relationships, albeit with minor discrepancies. For instance, the second relationship slightly overestimates at low stellar masses compared to observations. However, these observed relations are predominantly fits derived from data points concentrated above the low mass range (red diamonds). Moreover, the galaxy morphology could significantly influence the structure of these relationships. The panels suggest the existence of at least two distinct populations: one adhering to the observed relation (blue lines and red diamonds) and another aligning with the cloud of green triangles by \cite{reines2015}. The bulge-disk decomposition in galaxy selection, as highlighted by these authors, plays a pivotal role.

To further probe these populations, Figure \ref{fig:bh_mass2} depicts the BH versus stellar mass relationship (for ModA) by categorizing galaxies based on their $B/T$ ratio into disk galaxies ($B/T<0.3$, top panel), lenticulars ($0.3<B/T<0.7$, middle panel), and ellipticals ($B/T>0.7$, bottom panel). These plots vividly illustrate three distinct populations, each potentially adhering to its unique relation. Disk galaxies slightly overestimate compared to observations, whereas lenticulars align perfectly with the observed relation. Ellipticals bifurcate into two categories: one following a stringent correlation differing from the observed and disk galaxy relations, and the other mirroring the cloud of observed data by \cite{reines2015}. This underscores the significance of galaxy morphology in comprehending the BH versus stellar mass relationship.

In summary, both models adeptly capture the observed relations, marking a significant achievement for the SAM. More crucially, they also emulate the observed distinct populations in the BH-stellar mass relation, a remarkable feat considering that neither the BH-bulge mass, nor the BH-galaxy mass, relations were used for calibration. We will revisit this focal aspect in Section \ref{sec:overview}, juxtaposing our findings against existing observational knowledge.

\begin{figure*}
\centering
\includegraphics[width=0.95\textwidth]{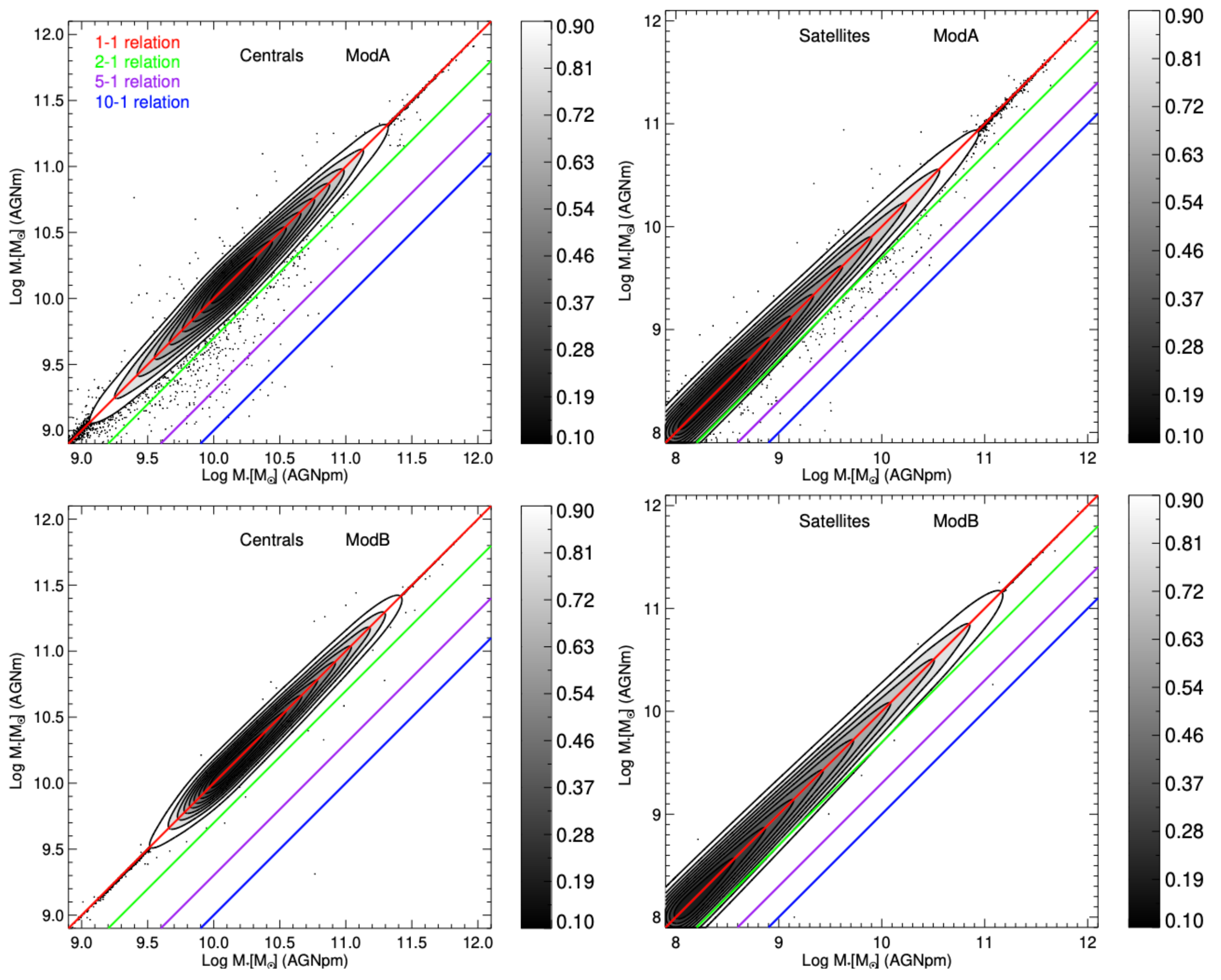}
\caption{In the top panels, we compare the relations between the stellar mass of central and satellite galaxies from ModA when both modes of the AGN feedback are active (x-axis), against that in which the positive mode is switched off (y-axis). The bottom panels show the same relations for ModB. A notable difference between the two models emerges from this comparison. In ModB, the change in stellar mass for central/satellite galaxies is relatively modest, with approximately 90\% of the data points falling within a 2-1 relation represented by the green line. In contrast, ModA exhibits more substantial variations in stellar mass, with several galaxies reaching and even surpassing a 10-1 relation. Excluding the most extreme cases reveals that the distributions of both models are nearly mirror images of the 1-1 relation. Nevertheless, only some central and satellite galaxies are actually more massive in the models where the positive AGN feedback is deactivated. For more details, refer to the text.}
\label{fig:BCG_mass}
\end{figure*}

\begin{figure*}
\begin{center}
\begin{tabular}{cc}
\includegraphics[width=0.9\textwidth]{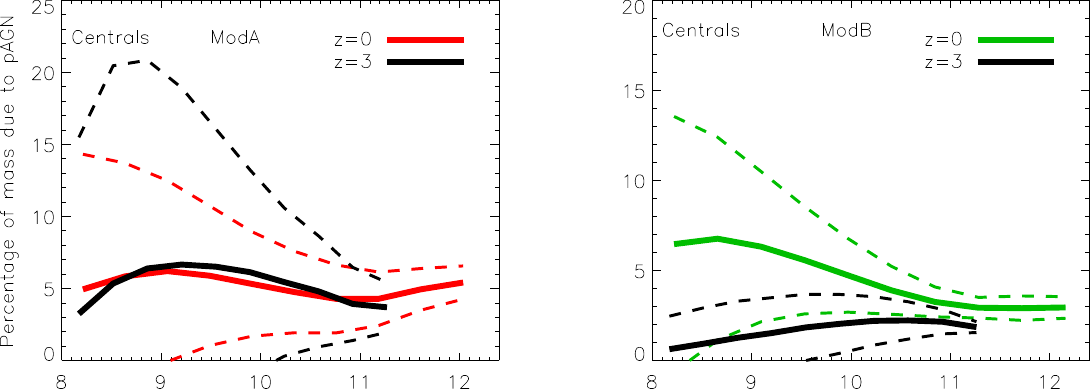} \\
\includegraphics[width=0.9\textwidth]{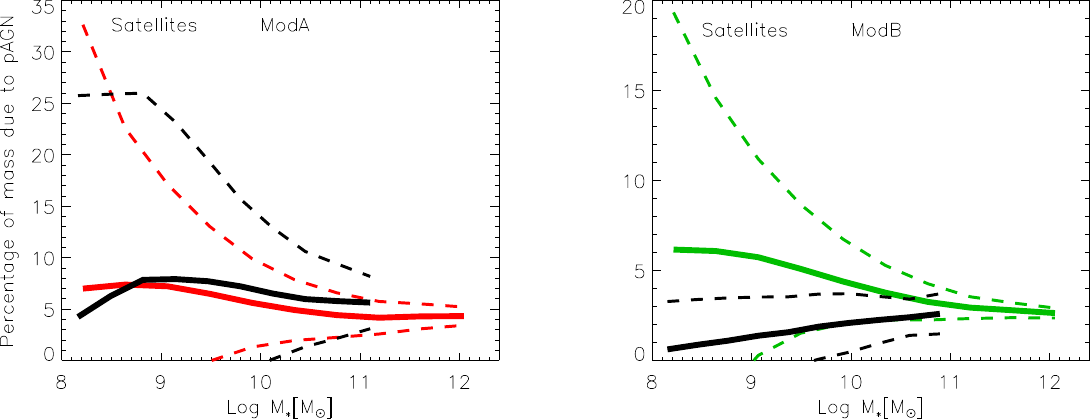} \\
\end{tabular}
\caption{In the top panels, we present the percentage of stellar mass contributed by the positive mode of the AGN feedback as a function of halo mass for central galaxies, while the bottom panels show the same for satellite galaxies. The predictions are made by ModA (red lines) and ModB (green lines) at two redshifts: $z=0$ (solid red/green lines) and $z=3$ (dash-dotted black lines). The plots account for all stellar mass attributable to the positive AGN feedback across all branches of the merger tree for each galaxy. Both models indicate minor differences between central and satellite galaxies and predict a decreasing percentage of stellar mass due to positive AGN feedback with increasing stellar mass, scatter included. This figure, combined with Figure \ref{fig:BCG_mass}, suggests that while galaxies can grow significantly when the positive AGN feedback is activated, this growth is largely the result of cumulative growth throughout their assembly histories. When considered individually, growth due to positive AGN feedback never exceeds an average of 7-8\%, even in dwarf galaxies. Interestingly, the trends between the models are reversed at $z=3$, with the positive mode in ModA being stronger on average than in ModB. This difference can be attributed to variations in SN feedback and reincorporation times applied in the models, which causes a stronger positive mode at lower redshifts in ModB. For further details, please refer to the text.}
\label{fig:perc_pAGN}
\end{center}
\end{figure*}

\subsection{AGN feedack and Galaxy Growth}\label{sec:agn_ggrowth}

In our concluding analysis, we delve into the pivotal role of the positive mode of AGN feedback, a cornerstone innovation in our SAM. Traditional AGN feedback primarily operates in its negative mode, which can prevent gas from cooling and consequently inhibit star formation. However, by integrating the positive mode into our SAM, we can explore its tangible impact on the final stellar mass of galaxies, differentiating between central and satellite galaxies.

To commence, we aim to quantify the stellar mass potentially acquired by central/satellite galaxies throughout their evolutionary history. We achieve this by comparing the stellar mass projected by ModA and ModB, considering both AGN feedback modes, against scenarios where only the negative AGN feedback mode is active. This comparative analysis is visualized in Figure \ref{fig:BCG_mass}, with ModA showcased in the top panels and ModB in the bottom panels. The contour distributions remain consistent with prior figures, while the colored lines denote various relations: 1-1 (red lines), 2-1 (green lines), 5-1 (purple lines), and 10-1 (blue lines).

The distributions within both models mirror the 1-1 relations, suggesting that while some centrals/satellites might gain stellar mass with the positive mode activated, others may lose it. This variability stems from intricate interactions of processes like satellite mergers with centrals, available reincorporated gas mass, and cooling hot gas mass. Broadly, a majority of centrals/satellites benefit from the positive mode, with some even surpassing the 10-1 relation by an order of magnitude or more. However, approximately 90\% of centrals and satellites are within the 2-1 relation, while a minor portion becomes more massive without the positive mode, albeit not significantly. Specifically, 25\% (31\%) of centrals (satellites) are more massive in ModA without the positive mode, while 5\% (18\%) of centrals (satellites) are more massive in ModB without the positive mode.

Moreover, ModA exhibits a more nuanced scenario compared to ModB. Indeed, in ModA there is a significant number of galaxies (both centrals and satellites) that not only exceed the 2-1 and 5-1 relations but also, in some cases, the 10-1 relation. For satellites, this growth remains more restrained. Yet, a majority of centrals/satellites remain within the 2-1 relation also in ModA, and only a fraction becomes more massive without the positive mode.

The disparity between the two models may be attributed to the varying power of the negative mode of AGN feedback. Specifically, the efficiency $\kappa_{\rm{AGN}}$ in ModB is five times lower than in ModA (refer to Table \ref{tab:parameters} for values), and consequently, the negative mode in ModB is generally less potent than in ModA. However, most of the difference can be attributed to the different power of the positive mode in the two models. According to Equation \ref{eqn:pAGN} and the values of the two parameters listed in Table \ref{tab:parameters}, the positive mode in ModA appears to be more
efficient. We will further discuss this point in Section \ref{sec:role_pAGN}.

The quantification of the impact of the positive AGN feedback through a straightforward comparison, as previously discussed, may be seen as overly simplistic. Since SAMs operate with merger trees, they inherently possess the capability to store the necessary data to provide a more nuanced understanding of the process when it is activated. By examining the merger tree of a particular galaxy, we can accurately determine the actual stellar mass formed during a burst triggered by the positive mode of AGN feedback. This analysis culminates in the insights presented in Figure \ref{fig:perc_pAGN}, focusing on both central and satellite galaxies.

When considering the growth of galaxies up to the present time, two pivotal observations emerge. Firstly, at $z=0$, ModA and ModB manifest similar percentages (somewhat lower in ModB for massive galaxies) for both centrals and satellites, although the scatter in ModA is generally higher than in ModB for satellites. Secondly, there is a discernible decrease in percentage with increasing stellar mass across both models and galaxy types, when the scatter is taken into account. This trend arises from Equation \ref{eqn:pAGN}, combined with the stronger negative mode in larger halos, which host larger BHs, limiting the available gas for star formation. Additional factors such as a galaxy's merging history and gas availability over time also contribute to this trend.

However, when examining growth solely up to $z=3$ (black lines), the dynamics shift. A reversal trend between ModA and ModB emerges, with ModA exhibiting greater efficiency than ModB at all stellar masses, being comparable to ModB's efficiency by $z=0$. This variation stems from the distinct SN feedback powers and reincorporation timescales between the two models. Specifically, ModA's SN feedback diminishes with redshift, while reincorporation is more consistent over time. Conversely, ModB predominantly reincorporates ejected gas at later stages, accounting for its increasing efficiency over time compared to ModA's decline.

In conclusion, focusing on ModA, galaxies (both centrals and satellites) with stellar masses less than $\log M_* \sim 9.0$ can generate a fraction greater than $\sim 20-25$\% (depending on the type of galaxy) from the positive AGN feedback burst implemented in {\small FEGA}. This percentage precipitously declines to $\sim 5$\% for more massive galaxies. Thus, the positive AGN feedback mode can harmoniously co-exist with its negative counterpart, and while its direct contribution to stellar mass formation may be modest in most cases, its influence on the broader evolutionary trajectory of galaxies is significant.

In the subsequent sections of this paper, we investigate deeper into the main findings, emphasizing the role of our positive AGN feedback implementation in the model. Despite the notable strides made by {\small FEGA} in both its iterations, it is not devoid of limitations. We will critically examine these limitations, propose potential avenues for refinement in future studies, and outline directions for future research.

\section{Discussion}
\label{sec:discussion}

The analysis in the previous section demonstrated that {\small FEGA} effectively reproduces the most significant scaling relations among galaxy properties. While {\small FEGA} shows promise, there remains potential for further improvements. In this section, we look deeper into our results, contextualizing them with existing observational evidence and predictions from various models. Section \ref{sec:overview} examines primary achievements of {\small FEGA} and their implications for galaxy formation. In Section \ref{sec:bhdiscussion}, we discuss the key relationships between the BH and its host galaxy. Lastly, Section \ref{sec:role_pAGN} hones in on the central focus of this study: the role of the positive mode of AGN feedback.

\subsection{Overview of the Galaxy Properties}\label{sec:overview}

Before delving into the main achievements of {\small FEGA}, it is important to highlight the key differences between ModA and ModB. As discussed in previous sections, ModA outperforms ModB due to varying prescriptions for SN feedback, reincorporation time of ejected material, and the strength of positive-mode AGN feedback. Additionally, some calibration parameters have different values between the two models. ModA features redshift-dependent SN feedback, stronger at higher redshifts and weakening over time, whereas ModB maintains consistently strong feedback. Consequently, ModB exhibits a longer reincorporation time compared to the smoother one in ModA. These differing prescriptions also influence the power of AGN feedback across both modes, including the positive mode, a topic we will explore further in Section \ref{sec:role_pAGN}.

One of the most crucial aspects that SAMs aim to capture is the number densities of galaxies across bins of stellar mass, i.e., the SMF. Reproducing the SMF at various redshifts is essential for understanding galaxy evolution and underpins many other scaling relations. Typically, models are calibrated based on the evolution of the observed SMF and are required to best replicate a myriad of observed scaling relations. While this approach is common among SAMs, the specific tools and methods for calibration can vary from one model to another.

{\small FEGA} effectively reproduces the overall evolution of the SMF while also predicting various other galaxy properties accurately. Consistent with recent SAMs, the high-mass end of {\small FEGA}'s SMF at the current epoch aligns well with observations from \cite{bernardi2018}. This characteristic of {\small FEGA}, and potentially other models, holds significant implications for the stellar-to-halo mass relation. Our analysis demonstrates that {\small FEGA} can accurately describe this relation across a wide range of halo masses, a fit further supported by the extended high-mass tail of the SMF observed by \cite{bernardi2018}.

The accuracy of matching the right stellar mass to the corresponding halo also impacts star formation. We have incorporated a revised star formation prescription that considers the role of pre-existing stars in the process of converting cold gas into stars. This model aligns with the observed extended KS relation from \cite{shi2011,shi2018}. According to this relation, our updated prescription assigns greater star-forming efficiency to larger galaxies, as detailed in Equation \ref{eqn:starform} and the parameter values provided in Table \ref{tab:parameters}. While this revised model has a minimal impact on the SMF evolution, it could be crucial for the SFR-mass relation. In Section \ref{sec:SFRmass}, we demonstrated that both versions of {\small FEGA} can replicate the observed main sequence from \cite{elbaz2007}. ModA performs notably better than ModB, exhibiting a more extended red sequence for larger galaxies and aligning more closely with the observed SSFR distribution and its bimodality—a critical aspect for theoretical models to capture accurately.

Despite its strengths, {\small FEGA} has some weaknesses, particularly in matching the observed red fraction of centrals and satellites as a function of stellar mass. While both versions show discrepancies, ModA performs better than ModB. {\small FEGA}, even in its improved ModA form, struggles to predict the observed percentages across most stellar mass bins, for satellite galaxies. For low-mass galaxies, it tends to over-predict the fraction for centrals and even more so for satellites. Conversely, it under-predicts at higher masses, with centrals showing much milder discrepancies than satellites. These mismatches could stem from issues related to gas stripping, reincorporation of ejected gas, or the cooling process. In contrast, the latest version of the GAEA model (\citealt{delucia2024}) aligns well with observed red fractions for both centrals and satellites. The authors attribute this improvement to their revised treatment of gas stripping and cold gas accretion onto BHs in \cite{xie2020} and \cite{fontanot2020}, respectively. This suggests potential avenues for enhancing {\small FEGA} in future iterations to better match observed fractions.

Considering the success of both ModA and ModB in replicating the expected galaxy fractions based on morphology (using the B/T in this study) as outlined by \cite{bluck2019}, it is worth examining the underlying prescription responsible for this performance. In Section \ref{sec:redfrac}, we argued that the improvement in our model stems from revising the disk instability prescription, introducing a new parameter, $\delta_{\rm{DI}}$, which brings the model predictions closer to observations. Our approach is not novel among SAMs; other studies, such as \cite{lagos2018} and \cite{izquierdo-villalba2019}, have also explored similar revisions to disk instability, often finding values of $\delta_{\rm{DI}}$ below the commonly assumed value of 1 in SAMs. As highlighted by H20, refining the bulge formation prescription in conjunction with the disk instability model could further enhance the predictive accuracy of galaxy morphologies. One potential enhancement might involve allowing some newly formed stars during the positive AGN feedback mode to be accreted into the bulge. Additionally, incorporating cold gas in the disk within a variable $\delta_{\rm{DI}}$ framework could also be beneficial. However, it must also be noted that the disk instability criterion (\citealt{efstathiou1982}), commonly adopted in SAMs, might not be an accurate description. In fact, as shown in \cite{romeo2023}, who tested the same criterion observationally, it fails in about 55\% of the cases, and it remains highly inaccurate even if the instability threshold (usually taken as $\sim$1) is adjusted up or down.

Like many other theoretical models, {\small FEGA} successfully describes the observed mass-metallicity relation, although it does not precisely match the trend at low stellar masses as observed by \cite{gallazzi2005}. However, this discrepancy is not unique to {\small FEGA} but rather a common feature of SAMs. Importantly, more recent observations from SDSS-DR7 (\citealt{zahid2017}) appear to support the predicted trend by our model and several others. Additionally, it is worth noting that the metal yield in ModA is half that of ModB (see $f_{\rm{metals}}$ in Table \ref{tab:parameters}), underscoring another distinction between the two versions of {\small FEGA} . Given the stronger SN feedback in ModB, as noted in H20, a higher metal yield is necessary to align with the mass-metallicity relation.

Concluding our overview of the key accomplishments of our SAM, we turn to its accuracy in replicating the stellar-to-halo mass (SHM) and stellar-to-halo mass ratio (SHMR) versus halo mass relations (\citealt{lin-mohr2004,behroozi2010,moster2010,reddick2013,behroozi2013,birrer2014,lu2015,moster2018,kravtsov2018,behroozi2019}, among others). As previously discussed, a focal aspect of these relations is their alignment with the high-mass end observed by \cite{bernardi2018} or within the range identified by these authors. This alignment is crucial for SAMs that use the evolution of the SMF for calibration. Matching this high-mass end within the observed scatter would also yield an accurate SHM and SHMR within the cluster range, which typically host the most luminous and largest galaxies at their centers. Deviations in the high-mass end of the SMF at $z=0$ could lead to a biased SHM, either underestimating or overestimating it, in the same halo mass range. Fortunately, our models, particularly ModA, align perfectly with the predicted SHM and SHMR relations.

All relations discussed in this section, except for the observed SMF at different redshifts, were not used in the calibration of {\small FEGA} and should be viewed as genuine predictions of the SAM. With this overview in mind, we now shift our focus to the central theme of this study: the impact of the positive AGN feedback mode implemented in {\small FEGA}.

\subsection{The BH-Galaxy Mass Scaling Relations}\label{sec:bhdiscussion}

A crucial aspect of galaxy formation in modeling AGN feedback is the relationship between the BH mass and that of the host galaxy. The BH can increase its mass during both quasar and radio modes of AGN feedback, with the latter mainly regulating gas cooling from relatively high redshifts to the present (\citealt{dimatteo2005}). In our AGN feedback scheme, the negative mode operates first, followed by the positive mode if any gas remains. While in reality, these modes likely operate concurrently, they both influence the growth of the galaxy and the BH. Thus, both modes potentially play a role in shaping the relationship between BH mass and stellar mass or bulge mass.

The investigation of the scaling relationship between BH mass and host galaxy properties has intrigued astrophysicists for decades. Early studies in the late 20th century suggested a possible link between BH and galaxy bulge masses. A seminal study by \cite{magorrian1998} revealed a strong correlation between a galaxy's bulge mass and its central BH mass, hinting at an evolutionary connection between bulge formation and BH growth. Subsequent research has further supported this co-evolution idea, indicating that galaxies with larger bulges typically host more massive BHs, and vice versa (\citealt{haring2004,ferrarese2005}).

Nevertheless, the precise nature of this scaling relation and the mechanisms driving it remain topics of ongoing research and discussion (e.g., \citealt{reines2015,bentz-manne2018,davies2018}). Some theories suggest that BH and bulge growth may be linked through processes such as galaxy mergers, gas accretion, and feedback (see, e.g., \citealt{kormendy-ho2013}). Recent studies have explored how factors like galaxy morphology, environment, and accretion processes influence the observed BH-bulge mass relation (see, \citealt{graham2007,mcconnell-ma2013,reines2015,davies2018} and \citealt{zhuang-ho2023} for a recent review).

In Section \ref{sec:bhmass}, we demonstrated that our models are capable to reproduce the BH mass as a function of both bulge and stellar mass. However, as noted by \cite{reines2015}, our models predict varying relationships depending on the galaxy population considered, specifically based on their B/T ratio. This suggests that the BH mass-host galaxy mass relationship is not universally consistent but varies with galaxy morphology. Nonetheless, the entire galaxy population, particularly for stellar masses above $\log M_* \sim 9.5$, can be represented by a single relationship, albeit less tightly constrained than typically assumed.

Furthermore, our SAM can currently predict the existence of BHs that are more massive than expected based on observed scaling relations, especially in dwarf galaxies, aligning well with recent numerical simulations (\citealt{weller2023}) and observations (e.g., \citealt{reines2015,ferre-mateu2021}). These BHs are also observed at high and very high redshifts, up to $z\sim 10$ (\citealt{fan2023,larson2023,maiolino2024,bogdan2024,mezcua2024}), suggesting that BH formation might not be universal. While the seeding model categories, light or heavy, are beyond the scope of this discussion, it is worth noting a key aspect: both types could potentially co-exist. Light seeding could lead to BH populations becoming more massive than expected based on their accretion history, while heavy seeding could naturally explain these BH populations already present at high redshift. This presents an intriguing challenge for theoretical models, including semi-analytics, which require high-resolution mass trees to be reliable.

\subsection{Role of Positive AGN Feedback}\label{sec:role_pAGN}

The idea to incorporate the positive mode of AGN feedback into our SAM arose from increasing observational evidence (see references above) suggesting that AGN can actually stimulate star formation rather than suppress it. This highlights the necessity for theoretical models of galaxy formation and evolution to consider the potential positive impacts of AGN feedback. The question should not be framed as \emph{''is the AGN feedback negative or positive?}", but rather as \emph{"can negative and positive modes of AGN feedback co-exist?}" Theoretical models must be adaptable enough to incorporate a star formation-triggering mode without compromising the overall depiction of the galaxy population; if anything, this addition should enhance it.

Motivated by this concept, we aimed to devise a method within the SAM to account for this observed phenomenon, which has been largely overlooked by previous models. Given the potentially non-linear nature of such a process, pinpointing it directly through a physical description—such as an equation or a system of equations linking relevant physical quantities—poses challenges. Initially, we hypothesized that the positive mode operates immediately following the negative mode, and any remaining cold gas could potentially be converted into stars. However, as previously noted, these two modes likely operate concurrently. The positive mode may be activated under specific conditions, possibly linked to the state of the hot gas undergoing cooling, in relation to the power of the jet/wind emanating from the central engine (see, e.g., \citealt{silk2009,silk2010,silk2013,silk2024} for an in-depth discussion on the variables that may be involved).

Our model, represented by Equation \ref{eqn:pAGN} and parameterized as shown in Table \ref{tab:parameters}, serves as an initial effort to isolate the positive mode. It should not be mistaken for a complete physical representation of AGN-triggered star formation. Rather, it is an attempt to distill a process for which we observe only the net effect, lacking full understanding of its underlying physics. Even with this limited scope, such a simplified description can enhance our understanding of galaxy formation and evolution while also addressing an observed phenomenon.

In the analysis presented in Section \ref{sec:agn_ggrowth}, we observed that both models anticipate a diminishing influence of positive AGN feedback as stellar mass increases. Specifically, for low-mass galaxies, both centrals and satellites with $\log M_* < 9.0$, both models attributes over 5\% of total stellar mass formation to this mode over the lifetime of the galaxies, and up to 25\% if considering the scatter. On the other hand, very massive galaxies do rely on positive AGN feedback for star formation on average with a similar amount, but they show quite a smaller scatter. In both models, galaxies with $\log M_* > 11.0$ generate just $\sim 5$\% or less of their stellar mass through this channel over their entire history, including all progenitors. Thus, the positive AGN feedback appears to have a limited impact on overall galaxy growth when considering star formation via this pathway. However, as we will explore further, deactivating this mode while maintaining other parameters, particularly the efficiency $\kappa_{\rm{AGN}}$ in the negative mode, reveals its cumulative effect on galaxy growth.

A crucial aspect of our analysis focuses on the role of positive AGN feedback across different redshifts. When considering the cumulative growth of galaxies up to $z=3$, a reversal in trend between the two models becomes evident. Specifically, accounting for growth up to the present, ModB's positive feedback mode exhibits roughly the same efficiency of ModA on average. However, when isolating growth at $z=3$ to gauge the power of the modes over redshift, ModA surpasses ModB at all masses (see Fig. \ref{fig:perc_pAGN}). This dynamic suggests that over the past approximately 11 billion years, ModB's positive feedback mode has become more potent, while ModA's has waned. In short, while the efficacy of ModA's positive AGN feedback diminishes over time, considering the scatters, it intensifies in ModB.

This contrasting behavior between the models primarily stems from their distinct approaches to SN feedback and the reincorporation of ejected material. In ModA, SN feedback weakens with decreasing redshift, and the reincorporation of ejected gas occurs gradually over time. Conversely, ModB maintains consistently strong SN feedback, devoid of any explicit redshift dependence. Ejected gas lingers in a reservoir for an extended period in this model, reincorporating only at very low redshifts due to an extended reincorporation time. This design is crucial for ModB to align with the evolution of the SMF, as indicated by H20 (though see also \citealt{henriques2015}). Consequently, a larger volume of gas is reincorporated at lower redshifts in ModB compared to ModA. If the negative mode fails to fully inhibit cooling, more gas becomes available for the positive mode, explaining the increasing power of the positive mode in ModB with decreasing redshift, while the trend reverses in ModA. Our ModA findings align qualitatively with earlier theories (e.g., \citealt{silk2024} and references therein) positing that positive AGN feedback is more potent at higher redshifts, succeeded by a highly effective negative mode. Howbeit, these are theoretical predictions, and current observational evidence (see, e.g., \citealt{gim2024} and references therein) remains insufficient to definitively exclude any model.

However, numerical simulations have increasingly emphasized the potential role of positive AGN feedback. A primary challenge in these studies is the resolution of the simulation, given that AGN feedback operates at a sub-grid level. Dedicated simulations are required to resolve spatially the jet (or wind) and the gas clump influenced by the feedback. On cosmological scales, achieving this level of resolution is unfeasible, necessitating the implementation of both negative and positive modes as sub-grid processes. Nonetheless, several theoretical studies offer insights into this area. For instance, \cite{gaibler2012} demonstrated that jets can stimulate star formation in disks, while \citealt{zubovas2017} (and references therein) highlighted the intricate interplay between AGN activity and star formation within host galaxies, where AGN feedback can manifest as both negative and positive simultaneously (see also the comprehensive review by \citealt{harrison2024} and references therein).

Despite the prevailing consensus regarding the influence of the negative mode within the scientific community, there is observational evidence supporting the role of the positive mode in triggering star formation (as cited above). Key questions revolve around the frequency of the positive mode and its implications for galaxy growth. While these concepts may seem straightforward, their investigation is far from trivial. Additional observational data, spanning various redshifts, are essential for accurately assessing the overall impact of the positive mode on galaxy growth, as previously mentioned. In the context of our SAM, both models suggest that the two modes may co-exist and exert influence at different intervals. The interplay between these modes is pivotal in our models. A direct comparison of the stellar masses of central and satellite galaxies between ModA and ModB, alongside the corresponding models with positive AGN feedback deactivated, reveals that, in most instances, galaxies exhibit greater mass when the positive mode is active. This outcome naturally stems from stars formed via this feedback channel.

Our current prescription for the positive mode of AGN feedback is rudimentary, and further refinement is warranted. Nonetheless, this study has demonstrated the feasibility of incorporating AGN-triggered star formation into SAMs. Similar methodologies could be adapted for simulations where AGN feedback operates as a sub-grid process. A prospective avenue for future research involves devising a more physically grounded prescription, building upon the one implemented here, to facilitate more accurate representations of the positive mode. Additionally, as reiterated throughout, our positive mode operates effectively only when the negative mode does not entirely inhibit cooling. Given the likely concurrent activity of both modes, our future plans include incorporating their simultaneous operation in the forthcoming, more advanced iteration of our SAM.

\section{Conclusions}
\label{sec:conclusion}

We have developed {\small FEGA} (Formation and Evolution of GAlaxies), a cutting-edge semi-analytic model that incorporates the most sophisticated prescriptions for the physical processes governing galaxy formation and evolution. Additionally, we introduced an enhanced prescription for star formation, which accounts for the extended KS relation. This revised approach factors in the star formation efficiency based on the stellar mass of a galaxy, thereby accounting for stars previously formed. Furthermore, we introduced a novel prescription detailing the positive mode of AGN feedback, which operates alongside its widely accepted negative counterpart. Two distinct versions of {\small FEGA} were constructed: (1) one featuring a mild SN feedback that is redshift-dependent, coupled with a gradual reincorporation time for ejected gas (ModA); (2) and another with robust SN feedback and an extended reincorporation time for ejected gas (ModB). Both models underwent independent calibration using an MCMC approach, leveraging the observed evolution of the SMF from $z=3$ to the present time.

In this study, our primary focus is to examine the influence of the positive mode of AGN feedback on the growth of the galaxy population. Having demonstrated that our SAM can reasonably replicate several key scaling relations among galaxy properties, we embarked on an exhaustive analysis. We began with the SMF evolution, which served as the calibration basis for our model, and subsequently delved into numerous scaling relations and other galaxy properties not employed in calibrating the two {\small FEGA} versions. In summary, both iterations of our SAM adeptly capture and elucidate the principal properties and scaling relations under scrutiny. Moreover, post-calibration, they accurately represent the SMF evolution up to the present epoch, a crucial criterion for any advanced SAM.

Regarding the general properties and scaling relations of galaxies, our primary findings can be summarized as follows:

\begin{itemize}
\item {\small FEGA} provides a robust description of the observed evolution of the SMF up to the present epoch. While this is not a direct prediction of the model, given its calibration on the SMF, a noteworthy result is that to replicate the stellar-to-halo mass relation across the entire halo mass spectrum, our SAM aligns more closely with the high-mass end of the SMF at $z=0$ as reported by \cite{bernardi2018}.

\item Both versions of the SAM effectively capture the SFR-mass relation. However, ModA exhibits a broader red sequence for larger stellar masses. This is also reflected in the SSFR distributions for galaxies within specific stellar mass ranges when compared to the observed distributions from SDSS-DR7.

\item The trend of the red galaxy fraction relative to stellar mass is well-represented by our models, showing an increasing fraction with increasing stellar mass. Nonetheless, both ModA and ModB deviate from the observed fractions at low and high stellar masses for both centrals and satellites. This discrepancy is less pronounced for central galaxies, with ModA exhibiting greater accuracy than ModB.

\item Our models adeptly reproduce the galaxy fraction relative to stellar mass when galaxies are categorized based on morphology, specifically the B/T ratio. The precision of our SAM in this regard is commendable, largely attributed to our revised disk instability prescription. Moreover, both models successfully replicate the observed mass-metallicity relation.

\item ModA and ModB, particularly the former, offer a compelling description of the stellar-to-halo and stellar/halo mass ratio against halo mass relations across the entire range of halo masses considered. ModA's alignment with predicted relations is particularly striking, further underscoring the merit of our SAM, given that these relations were not employed for model calibration.

\item Both models faithfully reproduce the observed BH-bulge/stellar mass relations with a high degree of accuracy. Moreover, the models predict distinct scaling relations when galaxies are segregated based on morphology (B/T criterion), aligning with observational findings. Specifically, while lenticular galaxies adhere to the commonly observed relation, disk galaxies follow a similar relationship but with elevated intercepts. Intriguingly, elliptical galaxies manifest both the observed relation with an augmented intercept and diminished slope, and a cluster of objects with BH masses lower than expected, as documented by \cite{reines2015}.
\end{itemize}

Our examination of the positive mode of the AGN feedback indicates that it can effectively co-exist with its negative counterpart. When the positive mode is deactivated in ModA and ModB, galaxies predominantly exhibit lower masses compared to their counterparts in the full models with active positive AGN feedback. While this outcome is a natural consequence of the positive AGN feedback, it is essential to note that not all differences can be solely attributed to it. Analyzing the growth in stellar mass attributed to the positive mode reveals that its efficacy is a decreasing function with stellar mass, being discretely significant primarily in dwarf galaxies, both centrals and satellites, in agreement with observations in the local Universe (e.g., \citealt{salome2015,schutte2022,gim2024}). Additionally, a notable disparity between the two models emerges: in ModA, the efficiency of the positive mode diminishes with redshift (aligning with prior theories), whereas in ModB, it ascends with redshift. We interpret these trends as inherent consequences of specific model implementations—SN feedback and reincorporation time. In ModB, the combined effects of these implementations render the AGN increasingly active over time as gas from the ejecta begins reincorporation, predominantly at lower redshifts due to extended reincorporation times. Conversely, in ModA, gas reincorporation progresses more uniformly over time, rendering the positive mode more effective at elevated redshifts.

{\small FEGA} has demonstrated its reliability as a SAM, effectively capturing key properties and scaling relations of galaxies. However, like all models, it is not flawless and presents avenues for enhancement. Firstly, forthcoming model iterations must address the observed fraction of red galaxies. We posit that this issue could be mitigated by refining gas stripping mechanisms as explored in \cite{delucia2024}, or through comprehensive revisions of potentially involved processes. Another promising direction could involve further calibrating the SAM using additional observations pertaining to the populations of red and blue galaxies, such as their independent SMFs or their redshift-specific fractions. Yet, considering insights from other SAMs, for instance, the {\small L-Galaxies} variant in \cite{henriques2020}, this approach might not offer a definitive solution to the issue.

In summary, we have analyzed the positive mode of the AGN feedback and demonstrated its compatibility with the negative mode. The comprehensive implementation of AGN feedback in {\small FEGA} stands to benefit from future refinements. This includes acknowledging the concurrent activity of both modes and, crucially, formulating a more realistic and physically grounded prescription for the positive mode, by incorporating a broader array of quantities pertinent to AGN activity.

\section*{Acknowledgements}
The authors thank the referee Darren Croton for his very constructive comments which helped to improve the manuscript.
E.C. and S.K.Y. acknowledge support from the Korean National Research Foundation (2020R1A2C3003769). E.C. and S.J. acknowledge support from the Korean National Research Foundation (RS-2023-00241934). All the authors are supported by the Korean National Research Foundation (2022R1A6A1A03053472). J.R. was supported by the KASI-Yonsei Postdoctoral Fellowship and by the Korea Astronomy and Space Science Institute under the R\&D program (Project No. 2023-1-830-00), supervised by the Ministry of Science and ICT.

\section*{Data Availability}
The current version of the code is not yet online, but an enhanced version will be publicly released once it is ready, accompanied by a notice in an upcoming publication. Meanwhile, code and galaxy catalogs used in this study can be obtained by contacting the corresponding author.

\bibliography{paper}{}
\bibliographystyle{aasjournal}

\end{document}